\documentclass{article}

\usepackage{PRIMEarxiv}

\usepackage[utf8]{inputenc} 
\usepackage[T1]{fontenc}    
\usepackage{hyperref}       
\usepackage{url}            
\usepackage{booktabs}       
\usepackage{amsfonts}       
\usepackage{nicefrac}       
\usepackage{microtype}      
\usepackage{lipsum}
\usepackage{fancyhdr}       
\usepackage{graphicx}       
\graphicspath{{media/}}     

\usepackage{algorithmic}
\usepackage{graphicx}
\usepackage{textcomp}
\usepackage{xcolor}
\usepackage{xspace}
\usepackage{multirow}
\usepackage{booktabs}
\usepackage{url}
\usepackage{hyperref}
\usepackage{wrapfig}
\usepackage{float} 
\usepackage{tikz}
\usepackage{lipsum}
\usepackage{paralist}
\usepackage{subcaption}
\usepackage{amsmath}
\usepackage{bm}
\usepackage{algorithm}
\usepackage{comment}
\usepackage{enumitem}

\newcommand{\rs}{\textit{RS}\xspace}
\newcommand{\sorl}{SORL\xspace}
\newcommand{\morl}{MORL\xspace}

\newcommand{\dis}{\textit{Dis}\xspace}
\newcommand{\ttc}{\textit{TTC}\xspace}
\newcommand{\rc}{\textit{RC}\xspace}
\newcommand{\jerk}{\textit{Jerk}\xspace}
\newcommand{\sd}{\textit{SD}\xspace}

\newcommand{\Atwelve}{\^{A}\textsubscript{12}\xspace}

\newcommand{\ov}{$OV$\xspace}
\newcommand{\sv}{$\#SV$\xspace}
\newcommand{\svs}{$SVS$\xspace}
\newcommand{\mv}{$\#MV$\xspace}
\newcommand{\mvs}{$MVS$\xspace}
\newcommand{\cc}{$\#C$\xspace}
\newcommand{\ub}{$\#UB$\xspace}
\newcommand{\ubd}{$UBD$\xspace}
\newcommand{\wbd}{$WBD$\xspace}
\newcommand{\scd}{$SCD$\xspace}

\title{Reinforcement Learning for Testing Interdependent Requirements in Autonomous Vehicles: An Empirical Study}

\author{
  Jiahui Wu \\
  Simula Research Laboratory and \\ University of Oslo \\
  Oslo, Norway \\
  \texttt{jiahui@simula.no} \\
  \And
  Chengjie Lu \\
  Simula Research Laboratory and \\ University of Oslo \\
  Oslo, Norway \\
  \texttt{chengjielu@simula.no} \\
  \And
  Aitor Arrieta \\
  Mondragon University \\
  Mondragon, Spain \\
  \texttt{aarrieta@mondragon.edu} \\
  \And
  Shaukat Ali \\
  Simula Research Laboratory \\
  Oslo, Norway \\
  \texttt{shaukat@simula.no} \\
}

\begin{document}
\maketitle

\begin{abstract}
Autonomous vehicles (AVs) make driving decisions without humans, making dependability assurance critical. Scenario-based testing is widely used to evaluate AVs under diverse conditions, with reinforcement learning (RL) generating test scenarios that identify violations of functional and safety requirements. Many requirements are interdependent and involve trade-offs, making it unclear whether single-objective RL (\sorl), which combines objectives into a single reward, can reliably reveal violations or whether multi-objective RL (\morl), which explicitly considers multiple objectives, is necessary. We present an empirical evaluation comparing \sorl and \morl for generating critical scenarios that simultaneously test interdependent requirements using an end-to-end AV controller and high-fidelity simulator. Results suggest that \morl and \sorl differ mainly in how violations occur, while showing comparable effectiveness in many cases. \morl tends to generate more requirement-violation scenarios, whereas \sorl produces higher-severity violations. Their relative performance also depends on specific objective combinations and, to a lesser extent, road conditions. Regarding diversity, \morl consistently covers a broader range of scenarios. Thus, \morl is preferable when scenario diversity and coverage are prioritized, whereas \sorl may better expose severe violations. Our empirical evaluation addresses a gap by systematically comparing \sorl and \morl, highlighting the importance of requirement dependencies in RL-based AV testing.
\end{abstract}


\section{Introduction}\label{sec:introduction}
\noindent
Autonomous vehicles (AVs) automatically navigate and make driving decisions without humans, aiming to achieve safe, efficient, and fully automated transportation. Thus, ensuring AVs' dependability before real-world deployment is essential but challenging due to the complexity of AVs and their driving environments. To this end, scenario-based AV testing~\cite{zhang2022finding,ding2023survey,stocco2022mind,stocco2023model,moghadam2024machine,neelofar2024identifying,crespo2024pafot} has emerged as a practical approach to testing AVs across diverse driving scenarios, helping identify functional and safety violations. However, given the complexity of AVs and the uncertainty of their driving environments, the number of possible driving scenarios for testing AVs is theoretically infinite, making it important to find critical scenarios that reveal requirement violations cost-effectively.

Search-based techniques~\cite{zhong2022neural,10645815,10234383,zhou2023specification} generate critical scenarios by exploring large search spaces guided by fitness functions. However, these approaches show limited effectiveness in handling runtime, sequential interactions when manipulating dynamic objects, as they typically do not adapt well to dynamic environmental changes during scenario generation.

Reinforcement learning (RL)~\cite{sutton2018reinforcement,lu2022learning,10172658,feng2023dense} has shown promise in dealing with these challenges by dynamically interacting with the environment to perform adaptive test scenario generation. 
However, the focus of RL-based testing has mainly been on evaluating individual requirements or testing multiple independent requirements. 
To test multiple \textit{independent} requirements, MORLOT~\cite{10172658} adapts single-objective RL and many-objective search to generate test suites to violate as many requirements as possible. However, in practice, many requirements are interdependent and must be evaluated simultaneously to ensure comprehensive testing. For example, violating a functional requirement, such as route completion, may compromise safety, as in a collision, highlighting the need for approaches that simultaneously evaluate multiple \textit{interdependent} requirements.

Currently, there is a lack of systematic empirical evidence on whether multiple interdependent requirements can be effectively tested using a weighted single-objective RL (\sorl), or whether multi-objective reinforcement learning (\morl), which inherently handles multiple objectives such as safety and performance by simultaneously optimizing multiple criteria through learned control policies, is necessary for adaptive testing of AVs in complex, dynamic, and uncertain environments~\cite{6918520,hayes2022practical}.

To address this gap, we present an empirical evaluation comparing \sorl with \morl for generating critical AV scenarios that simultaneously test multiple interdependent requirements. For \morl, we adopt Envelope Q-learning (EQL)~\cite{yang2019generalized}, a generalized algorithm that learns a single policy network to optimize multiple objectives by dynamically adjusting the agent’s preferences, i.e., the relative importance of each objective during training. For \sorl, we adopt a single-objective RL approach based on Deep Q-learning~\cite{mnih2015human}, where multiple objectives are combined into a single reward function through equal weighting. We evaluate \morl and \sorl using an end-to-end AV controller (Interfuser~\cite{shao2023safety}) and a high-fidelity simulator (CARLA~\cite{dosovitskiy2017carla}) across six roads covering diverse driving conditions.

Overall, our results indicate that \morl and \sorl achieve comparable effectiveness but differ in how requirement violations are manifested, with \morl generating a larger number of violation scenarios while \sorl tends to produce violations of higher severity. The relative effectiveness of the two approaches also varies across different objective combinations and road structures. In addition, \morl consistently demonstrates superior scenario diversity, generating a broader and more varied set of test scenarios. We also analyze the results in depth to identify the relative strengths and limitations of each approach, highlighting the necessity of explicitly considering requirement dependencies in RL-based AV testing. Based on this analysis, we derive insights into the relative suitability of \morl and \sorl for AV scenario generation, offering preliminary guidance for researchers and practitioners.

To summarize, the contributions of this paper are: 1) We present a systematic empirical evaluation comparing \sorl and \morl for generating critical AV scenarios that violate multiple interdependent requirements, highlighting the impact of requirement dependencies on testing effectiveness and scenario diversity; 2) We demonstrate how to formulate the AV testing problem as a multi-objective Markov decision process (MOMDP) and solve it using the EQL algorithm, guiding researchers to draw inspiration from this formulation for other software engineering problems; 3) We investigate the trade-offs and key factors influencing the performance of \sorl and \morl, and discuss their implications for selecting between the two algorithms in AV testing, providing empirical insights and preliminary guidance for researchers and practitioners.

\section{Background}\label{sec:background}

\subsection{Reinforcement Learning}\label{subsec:rl}
\noindent
Reinforcement learning (RL) learns an optimal policy for an agent to perform tasks through interacting with the environment to maximize cumulative rewards~\cite{sutton2018reinforcement}. Unlike supervised learning, which uses labeled data to train models, RL requires no prior knowledge and learns through trial and error. Specifically, in RL, an agent observes the current state of the environment, selects actions based on the policy, and receives a reward signal as feedback for its decision. A typical RL process can be formulated as a Markov Decision Process (MDP), which is a 4-tuple $<\mathcal{S}, \mathcal{A}, \mathcal{P}, \mathcal{R}>$ with state space $\mathcal{S}$ specifying the possible situations of the environment, action space $\mathcal{A}$ defining possible actions the agent can take, transition distribution $\mathcal{P}(s_{t+1}|s_t,a_t)$ specifying how the environment state changes in response to the agent's action, and reward function $\mathcal{R}(s_t,a_t)$ measuring the agent's performance of taking an action. In MDP, the agent interacts with the environment at each discrete time step. Specifically, at each time step $t$, the agent observes the current state of the environment as $s_t \in \mathcal{S}$ and then selects an action $a_t \in \mathcal{A}$ to perform based on $s_t$ and the behavioral policy. After the action is finished, the agent receives a reward $r_t \sim \mathcal{R}(s_t,a_t)$ as feedback to evaluate the agent's performance, and the environment moves into a new state $s_{t+1} = \mathcal{P}(s_{t+1}|s_t,a_t)$. RL aims to find an optimal policy $\pi^*(a_t|s_t)$, a mapping from state to probabilities of selecting actions, to maximize the expected cumulative reward over time. Q-learning~\cite{watkins1992q} is a classical RL algorithm that learns via an action-value function called Q-value function:
\begin{equation}
    Q^\pi(s_t, a_t) = \mathbb{E}_{\pi}[R_t|s_t, a_t].
\end{equation}
The Q-value of the state-action pair $(s_t, a_t)$ estimates the expected future reward by taking action $a_t$ in state $s_t$ with policy $\pi$. Q-value is updated based on the Bellman function:
\begin{equation}
    \label{equ:bellman_equation}
    Q^\pi(s_t, a_t) = \mathbb{E}_{\pi}[r_t + \gamma \max_{a_{t+1}}Q^\pi(s_{t+1}, a_{t+1})],
\end{equation}
where $\gamma$ is the discount factor determining how much future rewards contribute to immediate rewards.

Traditional RL algorithms, such as Q-learning, are designed to address tasks with a single long-term objective. However, many real-world tasks involve multiple criteria that must be optimized simultaneously. For example, when designing AVs, ensuring safety and passenger comfort is paramount. To this end, \morl, which learns optimal policies to handle two or more objectives, has become an important research area.

\subsection{Multi-Objective Reinforcement Learning}\label{subsec:morl}
\noindent
\morl aims to solve sequential decision-making problems where multiple objectives must be considered~\cite{roijers2013survey}. A \morl process can be formulated as an MOMDP, an extension of MDP. Unlike MDP, calculating a single scalar reward $r$, the reward function in MOMDP returns a vector $\mathbf{r}$ containing the rewards for each objective. The action-value function in \morl is a vectorized Q-function:
\begin{equation}
    \mathbf{Q}^\pi(s_t, a_t) = [Q_1^\pi(s_t, a_t),Q_2^\pi(s_t, a_t),...,Q_{NO}^\pi(s_t, a_t)]^T,
\end{equation}
where $NO$ is the number of objectives and for the $i_{th}$ objective, its corresponding action-value function is $Q_i^\pi(s_t, a_t)$, which is updated using the Bellman function~\ref{equ:bellman_equation}. \morl can be classified into two types based on how they find the optimal policy. The first class is single-policy \morl, aiming to find one optimal policy that represents the weights among the multiple objectives. The second class is multi-policy or Pareto-based \morl, which finds a set of policies that approximate the Pareto front. Each policy balances different objectives and no single policy can dominate another across all objectives.

In this paper, we employ a single-policy \morl called EQL~\cite{yang2019generalized} given its success in multi-objective RL tasks such as Games and Robot control~\cite{felten2024toolkit}. In EQL, an MOMDP is represented using a 6-tuple $<\mathcal{S}, \mathcal{A}, \mathcal{P}, \mathbf{R}, \Omega, f_{\Omega}>$, where $\mathcal{S}$, $\mathcal{A}$, and $\mathcal{P}$ are the same as in traditional MDP. Notably, $\mathbf{R}(s_t, a_t)$ represents a vector reward function, $\Omega$ is the space of weights of multi-objectives, and $f_{\Omega}$ is linear weight functions, each of which produces a scalar utility using the weight $\boldsymbol{\omega} \in \Omega$. The goal of EQL is to learn a single policy network that can be generalized across the entire weight space. One EQL implementation is based on Double Deep Q-learning (DDQN)~\cite{mnih2015human,van2016deep}, a classical single-objective RL that uses a deep neural network called Q-network with a target network to approximate and store Q-values and reduce overestimation bias. EQL extends the Q-network into a multiple-objective Q-network (MQ-network). Unlike Q-network, which takes only the state as input, MQ-network takes the concatenation of the state and the weight as input. 

As shown in Algorithm~\ref{alg:EQL}, EQL samples a weight vector $\boldsymbol{\omega}$ at the beginning of each episode to represent the relative importance of multiple objectives (line 7).
At each step, an action $a_t$ is selected either randomly or greedily according to the weighted sum of multi-objective Q-values produced by the MQ-network, which takes both the state $s_t$ and weight $\boldsymbol{\omega}$ as input (lines 9-10).
After executing the action $a_t$, the environment returns a vectorized reward $\boldsymbol{r}_t$ for all objectives and observes the next state $s_{t+1}$ (lines 11-13).
The transition $(s_t, a_t, \boldsymbol{r}_t, s_{t+1})$ is then stored in a replay buffer (line 14).
Once a minibatch of transitions is sampled from the replay buffer and a set of weights is drawn from the weight distribution, the best action and weight are selected to compute the target Q-values using the DDQN-style target network (lines 15-20).
These target Q-values are then used to calculate the training loss for updating the MQ-network parameters. The overall loss combines a standard vectorized Q-learning loss with an auxiliary homotopy loss, interpolated by a parameter $\lambda$ that gradually increases during training (lines 21-28).
Besides, the target network parameters are periodically updated to stabilize training (line 29).
These designs enable EQL to dynamically learn the relative importance of multiple objectives through adaptive weighting, thereby improving its ability to generalize across diverse problems.

\begin{algorithm}
\caption{Envelope Q-learning (EQL)}\label{alg:EQL}
\textbf{Input:}
\parbox[t]{0.9\linewidth}{
Environment $Env$, action space $\mathcal{A}$, objective functions $\mathcal{O}$, weight sampling distribution $\mathcal{D_\omega}$, path $p_\lambda$ for loss interpolation parameter $\lambda$ increasing from 0 to 1}

\begin{algorithmic}[1]
\STATE Initialize global environment $Env$
\STATE Initialize replay buffer $\mathcal{D}$, policy network $\boldsymbol{Q}(s,a,\boldsymbol{\omega};\theta)$, and $\lambda \gets 0$
\STATE Initialize target network $\boldsymbol{\hat{Q}}(s,a,\boldsymbol{\omega};\theta^-)$ with $\theta^- \gets \theta$
\FOR{each episode}
    \STATE Reset environment $Env$
    \STATE Observe initial state $s_0$
    \STATE Sample vectorized weight $\boldsymbol{\omega}$ from $\mathcal{D_\omega}$, where $len(\boldsymbol{\omega}) = |\mathcal{O}|$
    \FOR{each step $t$ of the episode}
        \STATE With probability $\epsilon$ select random action $a_t \in \mathcal{A}$
        \STATE otherwise select $a_t \gets \arg\max_a \boldsymbol{\omega}^\top \boldsymbol{Q}(s_t,a,\boldsymbol{\omega};\theta)$
        \STATE Execute action $a_t$ in $Env$
        \STATE Receive reward $r_t^o$ for each $o \in \mathcal{O}$ and observe next state $s_{t+1}$
        \STATE Compute vectorized reward $\boldsymbol{r}_t \gets [r_t^o]_{o \in \mathcal{O}}$
        \STATE Store transition $(s_t, a_t, \boldsymbol{r}_t, s_{t+1})$ in $\mathcal{D}$
        \STATE Sample $N$ transitions $(s_j, a_j, \boldsymbol{r}_j, s_{j+1})$ from $\mathcal{D}$
        \STATE Sample $N_\omega$ vectorized weights $W \gets \{\boldsymbol{\omega}_i\}_{i=1}^{N_\omega}$ from $\mathcal{D_\omega}$
        \STATE For all $j$ and $\boldsymbol{\omega}_i$, find best action and weight:
        \STATE $(a^*_j, \boldsymbol{\omega}^*_i) \gets \arg\max\nolimits_{a' \in \mathcal{A}, \boldsymbol{\omega}' \in W} \boldsymbol{\omega}_i^\top \boldsymbol{Q}(s_{j+1}, a', \boldsymbol{\omega}'; \theta)$
        \STATE Compute target using target network:
        \STATE $\displaystyle \boldsymbol{y}_{ij} \gets \begin{cases}
        \displaystyle \boldsymbol{r}_j, & \text{ if terminal } s_{j+1} \\
        \displaystyle \boldsymbol{r}_j + \gamma \boldsymbol{\hat{Q}}(s_{j+1}, a^*_j, \boldsymbol{\omega}^*_i; \theta^-), & \text{ otherwise}
        \end{cases}$
        \STATE Compute standard loss for transition $(s_j, a_j, \boldsymbol{r}_j, s_{j+1})$ and weight $\boldsymbol{\omega}_i$:
        \STATE $\displaystyle L_A(\theta) \gets loss(\boldsymbol{y}_{ij} - \boldsymbol{Q}(s_j, a_j, \boldsymbol{\omega}_i; \theta))$
        \STATE Compute auxiliary homotopy loss:
        \STATE $\displaystyle L_B(\theta) \gets loss(\boldsymbol{\omega}_i^\top \boldsymbol{y}_{ij} - \boldsymbol{\omega}_i^\top \boldsymbol{Q}(s_j, a_j, \boldsymbol{\omega}_i; \theta))$
        \STATE Combine losses:
        \STATE $\displaystyle L(\theta) \gets (1-\lambda) L_A(\theta) + \lambda L_B(\theta)$
        \STATE Perform a gradient descent step on $L(\theta)$ with respect to $\theta$
        \STATE Increase $\lambda$ along the path $p_\lambda$
        \STATE Periodically update target network parameters: $\theta^- \leftarrow \theta$
        \STATE $s_t \gets s_{t+1}$
        \IF{termination condition met}
            \STATE Exit current episode
        \ENDIF
    \ENDFOR
\ENDFOR
\end{algorithmic}
\end{algorithm}

\section{Approach}\label{sec:approach}
\subsection{\morl Strategy}\label{sec:overview}
\begin{figure}[!htbp]
    \centering
    \includegraphics[width=0.7\linewidth]{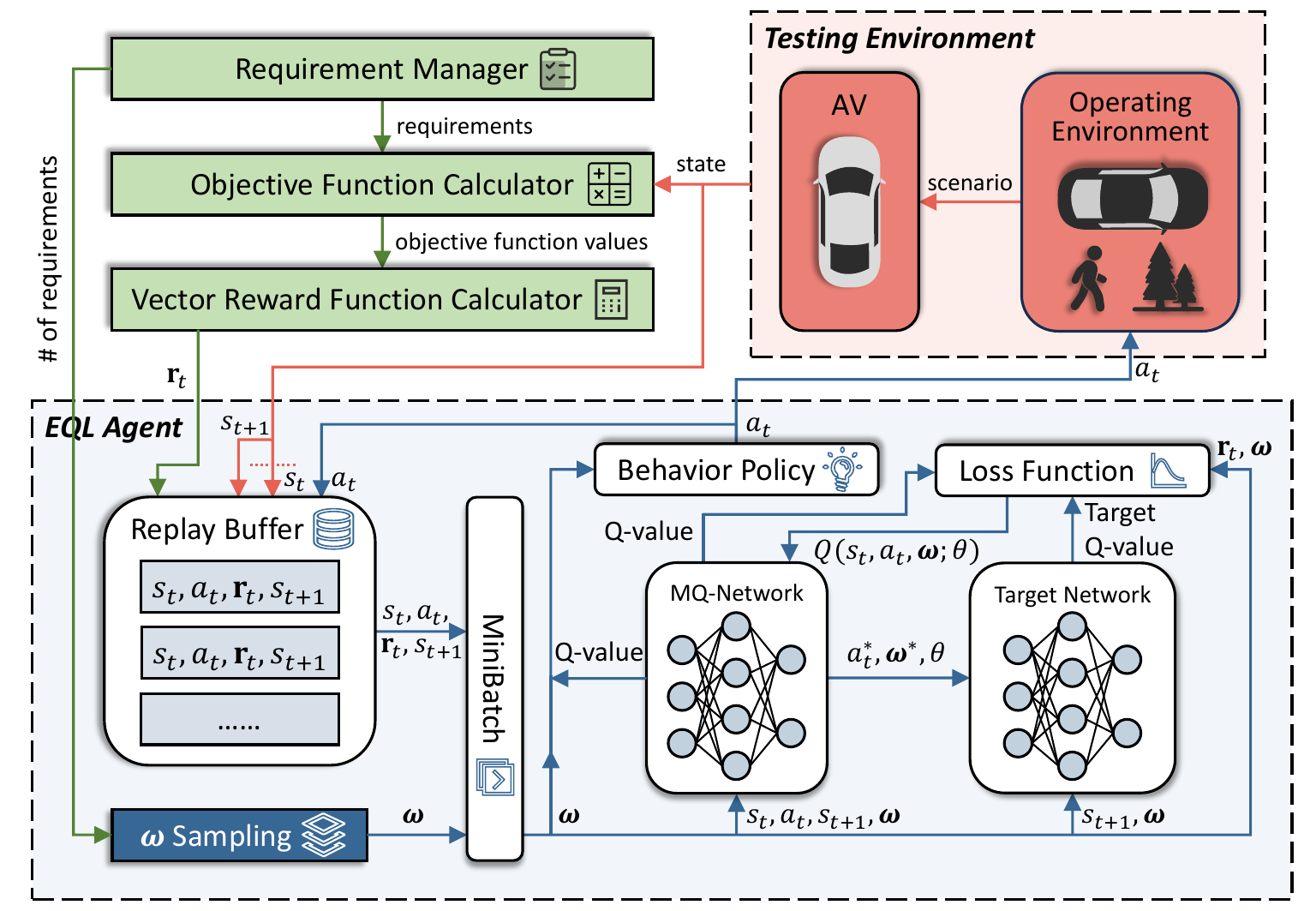}
    \caption{Overview of AV Testing with \morl}
    \label{fig:overview}
\end{figure}

\noindent Figure~\ref{fig:overview} shows an overview of how \morl generates critical scenarios to test multiple requirements simultaneously. Given a set of requirements, \morl employs EQL as the \morl solution to learn operating environment configurations to violate these requirements.
Such environment configuration characterizes a critical scenario, i.e., a test scenario in which the AV simultaneously violates multiple requirements. 

\morl builds a testing environment consisting of the AV and its operating environment. As Figure~\ref{fig:overview} shows, to configure the operating environment of the AV, at each time step $t$, the EQL agent observes the current state $s_{t}$ from the testing environment, which contains the state of both the AV and its operating environment. The agent then samples a vector of multi-objective weights $\boldsymbol{\omega}_t$ based on the number of requirements targeted for violation. Taking the concatenation of $s_t$ and $\boldsymbol{\omega}_t$ as input, the MQ-network computes the corresponding Q-values. Based on these Q-values and the weight vector $\boldsymbol{\omega}_t$, the behavior policy decides an action $a_{t}$ to configure the AV's operating environment.
The AV navigates in the newly configured environment for a specified period, during which the objective function calculator computes values in real-time based on the objective functions designed for each requirement violation, considering the state of the AV and the environment. These values are then fed into the vector reward function calculator for vector reward computation. 
After the AV reaches time step $t+1$, i.e., $a_t$ is finished, the EQL agent observes the next state $s_{t+1}$. The vector reward function calculator then calculates the corresponding vector reward $\mathbf{r}_{t}$ for the triplet $(s_{t}, a_{t}, \boldsymbol{\omega}_t)$ based on the objective function values during time step $t$. The agent then constructs a transition tuple $<s_{t}, a_{t}, \mathbf{r}_{t}, s_{t+1}>$ and stores it in a replay buffer using a specific replay mechanism. Notice that $\boldsymbol{\omega}_t$ is decoupled from the transition, which allows for sample-efficient learning by using the prioritized experience replay scheme~\cite{schaul2015prioritized}. Once a sufficient number of transitions have been collected in the replay memory, a minibatch sampled from the replay buffer, along with randomly sampled weights, are used to update the MQ-network. 
For each sampled transition and weight, the best action $a^*_t$ and weight $\boldsymbol{\omega}^*$ are determined by the online MQ-network. The target Q-value for this action-weight pair is then computed using a target network with an identical architecture to the MQ-network.
The loss is computed based on the target Q-value and the Q-value predicted by the MQ-network, and is used to update the network parameters.
The target network parameters are clones from the MQ-network every fixed number of steps.
Finally, \morl terminates an episode when the AV either completes its route, collides with other objects, or reaches the time limit (timeout).

\subsection{\sorl Strategy}\label{subsec:baseline}
\noindent
A single-objective RL approach with a weighted reward function (\sorl) is employed, which is used alongside \morl for comparison in the evaluation.
The corresponding scenario generation strategy is illustrated in Algorithm~\ref{alg:sorlw}.
Specifically, \sorl learns the optimal policy following a classical MDP process. At each time step, it randomly or greedily selects an action to perform in the environment based on the current state and the behavioral policy (lines 8-10). After the execution of the action, a reward is returned, and the environment enters a new state (lines 11-12). \sorl follows the same design of state space and action space as \morl (details will be provided later). However, unlike \morl, which adapts multiple reward functions, \sorl weights all reward functions into one reward function using equal weight, which is highlighted in line 12 of Algorithm~\ref{alg:sorlw}. The implementation of \sorl is based on Deep Q-learning~\cite{mnih2015human}, a classical RL algorithm that has demonstrated good performance in solving complicated sequential decision-making problems, such as robotics kinematics~\cite{phaniteja2017deep}, AV control~\cite{chen2020conditional}, and AV testing~\cite{lu2022learning}.

\begin{algorithm}
\caption{\sorl-based Scenario Generation Strategy}\label{alg:sorlw}
\textbf{Input:}
\parbox[t]{0.9\linewidth}{Environment $Env$, action space $\mathcal{A}$, objective functions $\mathcal{O}$}

\begin{algorithmic}[1]
\STATE Initialize global environment $Env$
\STATE Initialize replay buffer $\mathcal{D}$ and policy network $Q(s,a;\theta)$
\STATE Initialize target network $\hat{Q}(s,a;\theta^-)$ with $\theta^- \leftarrow \theta$
\FOR{each episode}
    \STATE Reset environment $Env$ and configure initial scenario
    \STATE Observe initial state $s_0$
    \FOR{each step $t$ of the episode}
        \STATE With probability $\epsilon$ select random action $a_t \in \mathcal{A}$
        \STATE otherwise select $a_t \gets \arg\max_a Q(s_t, a; \theta)$
        \STATE Execute action $a_t$ in $Env$
        \STATE Receive reward $r_t^o$ for each $o \in \mathcal{O}$ and observe next state $s_{t+1}$
        \STATE Compute weighted reward $r_t \gets \frac{1}{|\mathcal{O}|} \sum_{o \in \mathcal{O}} r_t^o$
        \STATE Store transition $(s_t, a_t, r_t, s_{t+1})$ in $\mathcal{D}$
        \STATE Sample minibatch of transitions $(s_j, a_j, r_j, s_{j+1})$ from $\mathcal{D}$
        \STATE Compute target for each transition:
        \STATE $y_j \gets \begin{cases}
        \displaystyle r_j, & \text{ if terminal } s_{j+1} \\
        \displaystyle r_j + \gamma \max\nolimits_{a'} \hat{Q}(s_{j+1}, a'; \theta^-), & \text{ otherwise}
        \end{cases}$
        \STATE Perform a gradient descent step on $loss(y_j - Q(s_j, a_j; \theta))$ with respect to $\theta$
        \STATE Periodically update target network parameters: $\theta^- \leftarrow \theta$
        \STATE $s_t \gets s_{t+1}$
        \IF{termination condition met}
            \STATE Exit current episode
        \ENDIF
    \ENDFOR
\ENDFOR
\end{algorithmic}
\end{algorithm}

\subsection{Problem Definition}\label{sec:problem}
\noindent AVs must meet a variety of safety and functional requirements that are not only for AVs but also for the environments in which they operate. Consequently, complex interdependencies exist among the requirements, making it ineffective to consider each requirement in isolation, e.g., for testing. Thus, to ensure the AV's safety and functionality, it is essential to validate safety and functional requirements simultaneously. 
To achieve this, scenario-based testing is employed to generate critical scenarios to violate various requirements, providing AV developers with a comprehensive testing solution.

Particularly, assume that given an AV, $Req = \{req_{1}, req_{2}, \dots, req_{n}\}$ is a set of its safety and functional requirements, and each $req_{i} \in Req$ represents a specific requirement that the AV must satisfy. For example, the AV should not collide with any object. 
Our goal in AV testing is to generate a critical scenario $sc$ using a function $g$ to simultaneously violate the requirements in $Req$.
Formally, this can be expressed as: 
\begin{equation}
    sc = g(AV, Req), 
    \label{eq:critical_scenario}
\end{equation}
where $AV$ is the AV under test and $Req$ is the set of requirements that need to be met.
To construct $sc$, along with the AV under test, the environment must be modeled. 
The environment $E$ can be defined as: 
\begin{equation}
    E = \{E_{static}, E_{dynamic}\}, 
\end{equation}
where $E_{static}$ represents a set of static objects within the scenario that remains static over time, e.g., trees and traffic signs, and $E_{dynamic}$ denotes dynamic objects in the scenario that may change over time, such as non-player character (NPC) vehicles and pedestrians. 
Static objects, defined as $O^{'}$, in $E_{static}$ are the objects that remain unchanged over time. Let $N$ be the number of static objects in the environment, where $E_{static}$ can be expressed as: 
\begin{equation}
    E_{static} = \{O_{1}^{'}, O_{2}^{'}, \dots, O_{N}^{'}\}. 
\end{equation}
Their configurations, denoted as $C^{O^{'}}$, can be represented using their states such as positions and rotations. 
In contrast, dynamic objects, defined as $O$, in $E_{dynamic}$ change over time, so their configurations $C^{O}$ must account for both their states and behaviors over time. 
Let $M$ be the number of dynamic objects in the environment, where $E_{dynamic}$ can be represented as: 
\begin{equation}
    E_{dynamic} = \{O_{1}, O_{2}, \dots, O_{M}\}. 
\end{equation}

To represent the dynamic changes in the environment over time, the states and behaviors of dynamic objects in $E_{dynamic}$ can be tracked across a sequence of time steps. Let $T$ be the total number of time steps and $t$ be the current time step, where $t \in \{1, 2, \dots, T\}$. Then $E_{dynamic}$ can be expressed as a time-sequenced set, described as: 
\begin{equation}
    E_{dynamic} = \{E_{dynamic}(t) | t \in \{1, 2, \dots, T\}\}. 
\end{equation}
Specifically, $E_{dynamic}(t)$ represents a set of dynamic objects at the time step $t$, which is defined as: 
\begin{equation}
    E_{dynamic}(t) = \{O_{i_{t}} | i \in \{1, 2, \dots, M\}\}, 
\end{equation}
where $O_{i_{t}}$ denotes the $i_{th}$ dynamic object in the environment at the current time step $t$.
Accordingly, over the entire time sequence, the $i_{th}$ dynamic object $O_{i}$ can be defined as:
\begin{equation}
    O_{i} = \{O_{i_{t}} | t \in \{1, 2, \dots, T\}\}. 
\end{equation}
Moreover, $O_{i_{t}}$ contains the set of configurations for the $i_{th}$ dynamic object that changes over instants within the time step $t$. 
Assuming each time step contains $m$ instants, we define $t_{j}$ as the $j_{th}$ instant within the time step $t$, and $O_{i_{t}}$ can be described as: 
\begin{equation}
    O_{i_{t}} = \{C_{t_{j}}^{O_{i_{t}}} | j \in \{1, 2, \dots, m\}\}, 
\end{equation}
where $C_{t_{j}}^{O_{i_{t}}}$ is the configuration of the dynamic object $O_{i_{t}}$ at instant $j$ within the time step $t$. 
Note that the dynamic environment of each time step (i.e., $E_{dynamic}(t)$) is influenced by the previous time step (i.e., $E_{dynamic}(t-1)$), with the configurations of all dynamic objects at the final moment of the prior time step (i.e., at the instant $m$ within the time step $t-1$, represented as $\{C_{{t-1}_{m}}^{O_{i_{t-1}}}\}^{M}_{i=1}$) continuously affecting the dynamic environment of the next time step.

\subsection{Requirements and Objective Functions}\label{sec:objective_function}
\noindent To quantify the degree or probability of violation for each requirement $req_{i}$ in the set $Req = \{req_{1}, req_{2}, \dots, req_{n}\}$ associated with an AV, we define a set of objective functions $H = \{h_{1}, h_{2}, \dots, h_{n}\}$. Each function $h_{i}$ measures the degree or probability of violation associated with the corresponding $req_{i} \in Req$ and can be expressed as:
\begin{equation}
    h_{i}(req_{i}) = \textit{Measure of violation for requirement } req_{i}, i \in \{1, 2, \dots, n\}. 
    \label{eq:general_objective_function}
\end{equation}
For instance, assuming one of the requirements for the AV is ``the AV should complete the route within a time budget", its corresponding objective function measures the percentage of the route completion. A smaller value of this objective function (i.e., a smaller route completion percentage) indicates a higher degree of violation. 
Thus, given a set of requirements $Req$, a critical scenario $sc$ (Equation~\ref{eq:critical_scenario}) is defined as the scenario that satisfies: 
\begin{equation}
    \forall req_i\in Req: sc \textit{ violates } req_i, i \in \{1, 2, \dots, n\}.
\end{equation}

\subsection{Formulating AV Testing as an MOMDP}\label{sec:MOMDP}
\noindent We model the \morl-based AV testing problem as an MOMDP as mentioned in Section~\ref{subsec:morl}, which is defined as a 6-tuple $<\mathcal{S}, \mathcal{A}, \mathcal{P}, \mathbf{R}, \Omega, f_{\Omega}>$. Specifically, $\mathcal{P}$ indicates the transition distribution, $\Omega$ describes the space of weights, and $f_{\Omega}$ expresses the weight functions. For AV testing, we adopt the default definitions for these components, while reformulating the state space $\mathcal{S}$, action space $\mathcal{A}$, and vector reward function $\mathbf{R}$ as follows. For comparison, \sorl corresponds to a standard MDP with the same state and action spaces, optimizing a single aggregated reward.

\subsubsection{State Space}
\noindent In \morl, the state space represents all the possible configurations of the environment that can be perceived and understood by the agent. Effective state encoding captures the critical information from the environment, enabling the agent to make appropriate decisions. For AV testing, a commonly used state encoding focuses on the AV and its operating environment~\cite{leurent2018survey}. Thus, the state space $\mathcal{S}$ can be formulated based on the configuration of the AV and its environment and can be represented as: 
\begin{equation}
    \mathcal{S} = \{C^{AV}, \{C^{O^{'}_{i}}\}^{N}_{i=1}, \{C^{O_{j}}\}^{M}_{j=1}\}, 
\end{equation}
where $N$ is the number of static objects, and $M$ is the number of dynamic objects. Notably, $C^{AV}$ means the configuration of the AV, $C^{O^{'}_{i}}$ is the configuration of the $i_{th}$ static object, and $C^{O_{j}}$ is the configuration of the $j_{th}$ dynamic object. 
At each time step $t$, the \morl agent can observe the specific state $\mathcal{S}(t)$ based on the formulated state space, which is denoted as: 
\begin{equation}
    \mathcal{S}(t) = \{C^{AV}_{t}, \{C^{O^{'}_{i}}_{t}\}^{N}_{i=1}, \{C^{O_{j}}_{t}\}^{M}_{j=1}\},
    \label{eq:specific_state}
\end{equation}
where $C^{AV}_{t}$, $C^{O^{'}_{i}}_{t}$, and $C^{O_{j}}_{t}$ denote the configuration of the AV, the $i_{th}$ static object, and the $j_{th}$ dynamic object at time step $t$, respectively.
Note that in the implementation, the actual state used by the \morl agent is obtained at a specific instant within each time step, corresponding to the definition provided in Section~\ref{sec:problem}.

\subsubsection{Action Space}
\noindent $\mathcal{A}$ encompasses all possible actions available to the \morl agent, defining the range of decisions it can make in a given state and directly determining the impact on the environment. At each time step $t$, the agent observes the current state $\mathcal{S}(t)$ (Equation~\ref{eq:specific_state}) and selects an appropriate action from the action space to transition the system to the next state $\mathcal{S}(t+1)$. This action involves manipulating objects in the AV testing environment, such as introducing dynamic objects or adjusting their configurations, thereby altering the environment to interact with the AV. These changes can create conditions that increase the likelihood of the AV making driving decisions that violate specified requirements.

\subsubsection{Vector Reward Function}
\noindent In \morl, $\mathbf{R}$ typically represents a vector reward function, with each reward function computing the reward value for a specific objective separately. After executing an action in the current state, the \morl agent receives a vector $\mathbf{r}$, which is an instance of $\mathbf{R}(t)$ at time step $t$. This reward vector contains rewards corresponding to multiple objectives and guides the agent in updating the weights for these objectives to balance their relative importance. 
For AV testing, $\mathbf{R}$ is formulated based on the objective functions (Equation~\ref{eq:general_objective_function}) defined in Section~\ref{sec:objective_function}. The objective functions $H = \{h_{1}, h_{2}, \dots, h_{n}\}$ assess the degree or probability of violations of requirements for each instant within every time step.
Since the direction of violation may differ among objectives (i.e., for some requirements, larger values of $h_i$ indicate more severe violations, while for others smaller values indicate worse performance), a monotonic transformation function $v_i(\cdot)$ is introduced for each $h_i$ to ensure that larger values consistently represent higher violation degrees. The function $v_i(\cdot)$ can be defined, for example, as a normalization or threshold-based mapping that converts the value of $h_i$ into a non-negative violation degree.
Let $t$ be the current time step and $\mathbf{R}(t)$ be the vector reward function at time step $t$. 
The vector reward function consists of $n$ individual reward functions, each corresponding to one objective, and can be written as:
\begin{equation}
    \mathbf{R}(t) = [\, r_1(t), r_2(t), \dots, r_n(t) \,],
    \label{eq:vector_reward_function}
\end{equation}
where each component reward function $r_i(t)$ is defined as:
\begin{equation}
    r_i(t) = \max_{j \in \{1, \dots, m\}} v_i(h_{i,t_j}),
    \label{eq:reward_function}
\end{equation}
where $h_{i,t_j}$ represents the value of the $i_{th}$ objective function at instant $j$ within time step $t$. 
These formulations ensure that $\mathbf{R}(t)$ captures, for each requirement, the maximum degree of violation observed within the time step, independent of the original scale or direction of the objective functions.
It also provides a clear one-to-one correspondence between each requirement, its objective function, and its reward function within the \morl framework.

\section{Experiment Design}\label{sec:design}

\subsection{Research Questions}\label{subsec:RQ}
The overall objective of our empirical evaluation is to systematically investigate the effectiveness and diversity of \sorl versus \morl in generating critical AV scenarios that violate multiple interdependent safety and functional requirements, to understand their relative strengths, limitations, and practical applicability for adaptive AV testing. Based on our overall objective, we aim to answer the following three Research Questions (RQs):

\begin{itemize}[left=0pt,label=\textbullet]

    \item \textit{\textbf{RQ0:} How do \sorl and \morl compare to a simple algorithm, i.e., Random Search (RS), in their effectiveness at generating critical AV scenarios that violate multiple interdependent requirements?} This RQ serves as a sanity check to determine whether \sorl and \morl are needed, or if a simple algorithm, such as RS, is sufficient, which would indicate that the problem we are addressing is relatively simple. To assess this, we conducted a pilot study comparing \sorl and \morl with \rs. The results show that \sorl and \morl significantly outperform \rs in terms of effectiveness. Therefore, \rs is excluded from the comparison in the main experiments. Detailed results of this pilot study are available in our online repository~\cite{MORLrepo}.
    
    \item \textit{\textbf{RQ1:} How do \sorl and \morl compare in terms of their overall ability to generate critical AV scenarios that violate multiple interdependent requirements across different objective combinations and road structures?} This RQ aims to compare the overall relative performance of \sorl and \morl in terms of both effectiveness and diversity in generating multi-requirement violation scenarios among all the objective combinations and road structures.

    \item \textit{\textbf{RQ2:} How do different objective combinations affect the comparative effectiveness and diversity of \sorl and \morl in critical AV scenario generation involving violations of multiple interdependent requirements?} This RQ aims to investigate how varying combinations of optimization objectives influence the comparative performance of \sorl and \morl under the generation of critical AV scenarios with multiple requirement violations.

    \item \textit{\textbf{RQ3:} How does the relative performance of \sorl and \morl differ across different road structures in multi-requirement violation scenario generation?} This RQ aims to examine whether the relative performance of \sorl and \morl changes depending on the different road topologies and layouts.

\end{itemize}

\subsection{Subject System and Simulator}\label{subsec:simulator}
\noindent We employ Interfuser~\cite{shao2023safety} as the subject system, which is an advanced end-to-end AV controller on the CARLA leaderboard. Interfuser is a safety-enhanced framework based on the Transformer architecture~\cite{vaswani2017attention} for comprehensive scene understanding and safe autonomous driving. It employs a multimodal sensor fusion transformer encoder to integrate sensor data from multiple RGB cameras and a LiDAR sensor. It uses a transformer decoder to generate driving actions and interpretable intermediate features. Finally, a safety controller processes the interpretable intermediate features to refine and enhance the safety of the driving actions. Interfuser has been evaluated in extensive driving scenarios, from urban environments to highways, demonstrating its outstanding safe driving performance. 
Regarding the simulator, we adopt CARLA~\cite{dosovitskiy2017carla}, a high-fidelity open-source autonomous driving simulator. CARLA provides an extensive list of digital assets, such as vehicles, sensors, and high-definition maps, to support AV development, training, and validation.
In the evaluation of \sorl and \morl, we use CARLA-0.9.10.1 and its default settings for Interfuser. We also chose the default vehicle used by Interfuser, which is a Tesla Model 3.

\subsection{Experiment Setup}\label{subsec:setup}

\subsubsection{Instantiation of Requirements into Objective Functions}\label{subsubsec: requirements_objective_functions}
\noindent While our approach is generic and applicable to a wide range of AV requirements, in our evaluation, we consider five representative and interdependent requirements, which include both safety-critical and functional aspects. The proposed approach is not limited to these requirements, and other AV requirements can be easily incorporated by defining corresponding objective functions. For the selected requirements, we formulate objective functions for AV testing by converting the requirements into mathematical expressions that quantify how close the RL algorithm is to violating them, rather than relying solely on binary requirement-violation information. The following are the safety and functional requirements we selected for our study:

\begin{itemize}
\item \textit{R1: The AV should keep a safe distance from surrounding obstacles to avoid collisions.}
\item \textit{R2: The AV should maintain a sufficient time to collision with nearby objects to prevent imminent crashes.}
\item \textit{R3: The AV should complete its route and reach the destination within the specified time budget.}
\item \textit{R4: The AV should drive smoothly with stable acceleration to ensure passenger comfort.}
\item \textit{R5: The AV should maintain a reasonable speed range consistent with surrounding traffic.}
\end{itemize}

\textit{R1} emphasizes that the AV must maintain a safe distance from surrounding static and dynamic obstacles to avoid physical collisions. This requirement focuses on spatial separation to ensure the safety of AV and prevent traffic accidents in the driving environment~\cite{tuncali2019requirements}. 
To measure the violation of this requirement, we apply \textit{Distance (Dis)}~\cite{westhofen2021criticality} as the corresponding objective function.
\dis calculates the minimum Euclidean distance between the AV and the other objects in the environment, which is defined as:
\begin{equation}
Dis = \min_{i} \| p_{AV} - p_{obj_i} \|_2  \label{eq:dis},
\end{equation}
where $p_{AV}$ and $p_{obj_i}$ represent the positions of the AV and the $i_{th}$ object, respectively. The value of \dis ranges from 0 to positive infinity. A smaller \dis indicates a higher likelihood of violating \textit{R1}. When \dis falls below a predefined minimum safety distance (e.g., 5 m), it is considered a complete violation of \textit{R1}. From the value of \dis, it intuitively and directly reflects the safety level of the AV’s operation, providing a clear indication of potential safety risks in the driving environment.

\textit{R2} highlights that the AV should maintain a sufficient time to collision with surrounding obstacles, reflecting a temporal safety constraint. Unlike \textit{R1}, which focuses on spatial separation, this requirement ensures that the AV has enough time to react and avoid potential collisions with both static and dynamic objects~\cite{vogel2003comparison}.
To quantify the violation of this requirement, we use \textit{Time to Collision (TTC)}~\cite{minderhoud2001extended} as an objective function.
\ttc represents the minimum remaining time before the AV collides with all other objects, assuming both maintain their current collision routes and velocity differences, formulated as:
\begin{equation}
TTC = \min_{i} \frac{d_{rel,i}}{v_{rel,i}}  \label{eq:ttc},
\end{equation}
where $d_{rel,i}$ means the relative distance between the AV and the $i_{th}$ object in the environment, and $v_{rel,i}$ calculates the relative velocity of the AV to the $i_{th}$ object. 
The value of \ttc ranges from 0 to positive infinity. 
A lower \ttc indicates a higher possibility of violating \textit{R2}. 
Specifically, we can preset a safety threshold for the \ttc value: when \ttc is smaller than this threshold, it indicates a complete violation of \textit{R2}.
We can perform real-time calculations on \ttc, offering insights into the likelihood of potential collisions from the estimated time remaining before impact, thereby enhancing the testing of AV driving safety. 
Note that, for simplicity, we provide a general formulation for \ttc; however, in our implementation, we compute the actual \ttc based on the specific position and velocity vectors of the AV and objects within the scenario.

The \textit{R3} requirement demonstrates our focus on the functional aspects of the AV. It requires the AV to complete the planned route and reach its destination within a specified time budget, which involves real-time environmental awareness and path planning to dynamically adapt to complex scenario changes, verifying the AV's driving reliability~\cite{schwarting2018planning}. To assess this requirement, we employ \textit{Route Completion (RC)}~\cite{carlaleaderboard} as an objective function, measuring the percentage of route successfully completed by the AV. \rc is defined as follows:
\begin{equation}
RC = \frac{d_{driven}}{D_{total}} \times 100\%  \label{eq:rc},
\end{equation}
where $d_{driven}$ measures the actual distance traveled by the AV along a given route, and $D_{total}$ denotes the total length of the entire test route. The value of RC ranges from 0\% to 100\%, with 100\% indicating that the AV has successfully completed the specified route. 
By measuring RC, we can assess the AV's execution efficiency and the stability and reliability of its global or local path planning in dynamic driving conditions. 
Thus, for AV testing, we expect a lower RC, which indicates a higher likelihood of the violation of \textit{R2}. 
Note that if RC stays at 0\% continuously, it indicates that the AV is either unresponsive or remains stopped, which should be avoided in the implementation as it may result in an ineffective test.

\textit{R4} also reflects another functional requirement in AV testing, although from the comfort perspective. This requirement expects the AV to maintain a relatively stable acceleration during driving to enhance passenger comfort. Satisfying this requirement contributes to improving passengers’ acceptance and satisfaction with autonomous driving technology~\cite{de2023standards}. To quantify the degree of violation of this requirement, we use \jerk~\cite{feng2017can} as the corresponding objective function, defined as follows:
\begin{equation}
Jerk = \frac{a_{t+1} - a_t}{\Delta t}  \label{eq:jerk},
\end{equation}
where $a_{t+1}$ and $a_t$ denote the accelerations of the AV at time $t+1$ and $t$, respectively, and $\Delta t$ represents the time interval between them. The value of \jerk ranges from 0 to positive infinity. In AV testing, a higher \jerk value indicates a greater degree of violation of \textit{R4}. This objective function highlights the importance of acceleration variations in vehicle dynamic control and is crucial for testing and developing motion planning algorithms that aim to maximize ride comfort. Note that, in implementation, $\Delta t$ is set to a fixed value to ensure fair computation, and the corresponding acceleration data are used to calculate \jerk accordingly.

\textit{R5} emphasizes that the AV should adjust its speed based on the surrounding traffic flow to maintain a reasonable speed range. For example, when the AV is in dense traffic, vehicles tend to slow down to avoid collisions, while in sparse traffic conditions, they are likely to increase speed for greater efficiency. Since the AV’s speed is closely related to traffic safety, maintaining a speed consistent with surrounding vehicles helps reduce collision risks~\cite{chen2023safe}. Therefore, we adopt \textit{Speed Difference (SD)}~\cite{carlaleaderboard} as the objective function to evaluate whether the AV’s speed remains within a similar range to that of nearby vehicles.
Considering the specific road conditions, we set an upper speed limit for the surrounding vehicles, denoted as $v_{max}$. Let $v_i$ represent the speed of the $i_{th}$ surrounding vehicle, and $N$ represent the total number of surrounding vehicles. The mean speed of the surrounding vehicles is then formulated as:
\begin{equation}
v_{mean} = \frac{1}{N} \sum_{i=1}^{N} \min(v_i, v_{max}),
\end{equation}
To account for variations in traffic dynamics, a proportional coefficient $R$ is introduced to relax the criterion for the minimum reasonable speed of the AV. Accordingly, the minimum reasonable speed $v_{min,r}$ that the AV should maintain to align with the surrounding traffic is given by $v_{min,r} = R \cdot v_{mean}$, and the maximum reasonable speed is set as $v_{max,r} = v_{max}$. Based on these definitions, the \sd of the AV from the surrounding traffic can be defined as:
\begin{equation}
SD = \begin{cases}
v_{min,r} - v_{AV}, & \text{if } v_{AV} < v_{min,r} \\
v_{AV} - v_{max,r}, & \text{if } v_{AV} > v_{max,r} \\
0, & \text{otherwise}
\end{cases},
\label{eq:sd}
\end{equation}
where $v_{AV}$ denotes the current speed of the AV. Theoretically, the value of \sd ranges from $[0, +\infty)$. A larger \sd indicates a higher degree of violation of \textit{R5} by the AV. Specifically, when the value of this objective function is nonzero, it implies that the AV has completely violated \textit{R5}. \sd provides an indirect measure of an AV’s driving safety and risk-taking behavior, offering a novel perspective for AV testing and development.

\subsubsection{Configurable Environment Parameters}\label{subsubsec: parameters}
\noindent When applying scenario-based testing to AVs, some degree of abstraction is needed, as considering all environmental factors that occur in reality would require enormous computing power, making our approach non-scalable. 
We therefore prioritize, for our study, the configuration and analysis of dynamic objects, i.e., NPC vehicles, although it could easily be extended to factors that could be of interest to AV developers.

\begin{figure}
    \centering
    \includegraphics[width=0.8\linewidth]{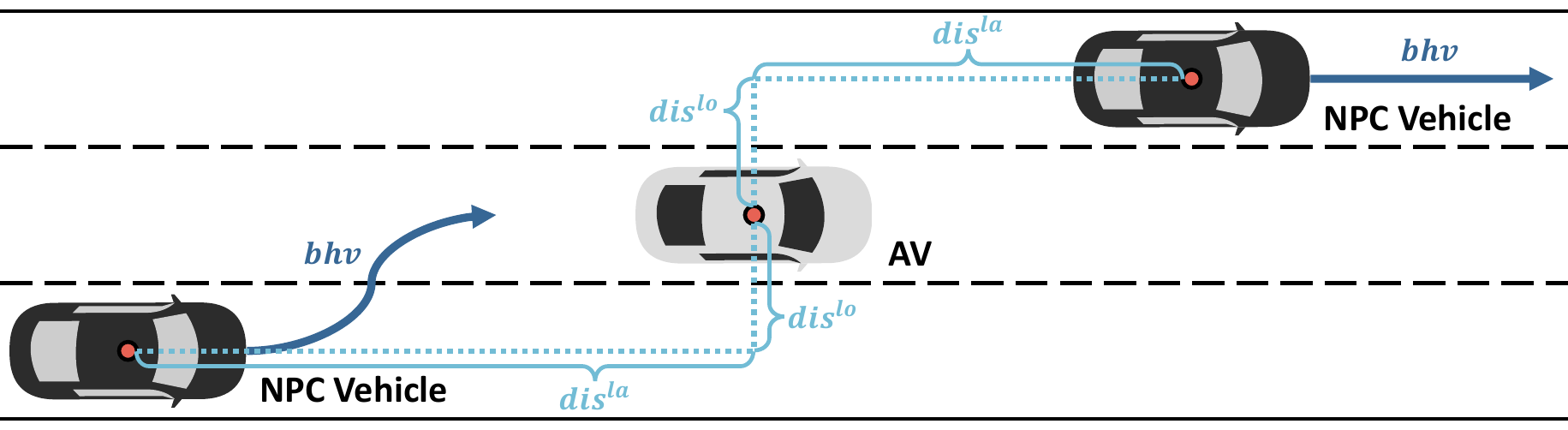}
    \caption{NPC Vehicle Configurable Parameter Examples}
    \label{fig:conf_env_parameter}
\end{figure}

The configurable parameters of an NPC vehicle can be represented by a triplet: $(dis^{la}, dis^{lo}, bhv)$, where $dis^{la}$ and $dis^{lo}$ mean the lateral and longitudinal distances between the NPC vehicle and the AV, respectively. Meanwhile, $bhv$ indicates the driving behavior of the NPC vehicle. An example of the configurable parameters for NPC vehicles is illustrated in Figure~\ref{fig:conf_env_parameter}.
In our implementation, to simplify the problem and enhance the experimental efficiency, based on the results of our pilot study and supported by findings in related research~\cite{lu2024epitester}, we sample the value of $dis^{la}$ within the ranges of $[-10, -5]$ and $[10, 35]$, and $dis^{lo}$ within the range of $[-25, 20]$. 
The value ranges of $dis^{la}$ and $dis^{lo}$ are determined by two main considerations: first, ensuring that the initial positions of the NPC vehicles maintain a safe distance from the AV; and second, preventing the NPC vehicles from being too far away from the AV, which would make the test ineffective.
Since $dis^{la}$ cannot take a value of 0 or close to 0, allowing $dis^{lo}$ to take 0 or near 0 values does not pose any safety risk to the AV.
Note that the reasonable ranges of $dis^{la}$ and $dis^{lo}$ depend on the specific road structure. When the road configuration changes, these ranges should be adjusted accordingly. The specific road structures considered in our implementation will be described in detail in Section~\ref{subsubsec:roads}, where the valid ranges of $dis^{la}$ and $dis^{lo}$ for each road type will correspond to a subset of the overall range.
The behavior $bhv$ of an NPC vehicle can be one of six options: (1) lane keeping, (2) right lane change, (3) left lane change, (4) acceleration, (5) deceleration, and (6) emergency braking, which are selected to represent the most common and safety-critical longitudinal and lateral driving behaviors in driving scenarios~\cite{galvao2024pedestrian}.
For example, if $(dis^{la}, dis^{lo}, bhv)$ is set to $(-10, 3.5, \textit{left lane change})$, it means that the NPC vehicle will be generated 10 meters directly behind the AV and 3.5 meters to the right, with its behavior set to prepare for a left lane change. 
Note that the travel direction of the generated NPC vehicle is set to align with the designated direction of the road it is on. 
Furthermore, in the actual experiments, the values of $dis^{la}$ and $dis^{lo}$ for NPC vehicles are configured only during the initialization phase. Thereafter, these distance parameters are dynamically updated according to the initially assigned $bhv$ and the subsequently selected $bhv$.

Moreover, to ensure the realism of the generated NPC vehicles in the scenario, we apply specific realistic constraints. As previously mentioned, NPC vehicles are placed at a safe distance from the AV. Besides, newly generated NPC vehicles are also positioned at a safe distance from other NPC vehicles already present in the scenario. All NPC vehicles are confined within the boundaries of the map on which the AV is traveling and are generated on appropriate roads (e.g., lanes) specified by the simulator, preventing unrealistic placements (e.g., on top of a tree). 
Moreover, the driving direction of generated NPC vehicles is aligned with the designated road direction, and once initialized, these NPC vehicles are controlled by the simulator's default autopilot algorithm. 
Therefore, apart from the initialization phase, where the positions of NPC vehicles are configured, the subsequent position changes are jointly controlled by the default autopilot algorithm and the predefined or subsequently selected driving behaviors, ensuring the realism of NPC vehicle movements.
For configured behaviors such as lane changes, corresponding commands are sent to the NPC vehicle, while the actual control, e.g., lane-changing control, is executed by its default algorithm.
Acceleration, deceleration, and emergency braking behaviors are implemented by adjusting the throttle and brake inputs of the NPC vehicle, ensuring that these transitions occur in a smooth and continuous manner consistent with real-world dynamics.
In addition, to prevent the NPC vehicle from exceeding the speed limit, a speed control mechanism is introduced: when the vehicle’s speed surpasses the limit, the throttle and brake are adjusted to reduce the speed and keep it within a safe range.
These constraints ensure the realistic generation and dynamic control of NPC vehicles.

\subsubsection{MOMDP Instantiation}\label{subsubsec:momdp}
\noindent 
Previous studies have indicated that using only two NPC vehicles is sufficient to construct testing scenarios that effectively evaluate the AV~\cite{ji2025autonomous,10.1145/3540250.3549100,deng2025target,tang2024legend}. Therefore, in this study, we initialize two NPC vehicles at the beginning of each scenario and dynamically assign appropriate driving behaviors to them at every time step, enabling a continuous and realistic simulation of traffic environments. The MOMDP formulation is accordingly instantiated for one AV and two NPC vehicles. Based on our instantiation strategy, the MOMDP model can be further extended to include additional NPC vehicles, offering a flexible and scalable testing framework for researchers and developers.
For comparison, the standard MDP formulation for \sorl is instantiated similarly to the MOMDP, sharing the same state, action, and transition dynamics, and differing only in the reward function instantiation via aggregation of multiple rewards into a single scalar reward.

\textbf{State Space.} 
Leveraging the parameters available from the CARLA simulator, we define the specific state for the AV at each time step as a quintuple $<pos_{AV}, rot_{AV}, vel_{AV}, acc_{AV}, anv_{AV}>$. 
Each element of the state represents the AV's current position, rotation, velocity, acceleration, and angular velocity, respectively.
Specifically, each element consists of three components. For example, the position of the AV can be represented as $pos_{AV} = (x_{AV}, y_{AV}, z_{AV})$. However, in the practical simulation process, certain components remain constant. For instance, the vertical position $z_{AV}$ of the AV typically remains zero, as the vehicle drives on a planar road surface, excluding cases such as uphill, downhill, or other unrealistic situations (e.g., the vehicle flying off the ground). To ensure parameter validity, such invariant components are removed. Consequently, the AV state is defined as $pos_{AV} = (x_{AV}, y_{AV})$, $rot_{AV} = (yaw_{AV})$, $vel_{AV} = (x_{AV}, y_{AV})$, $acc_{AV} = (x_{AV}, y_{AV})$, and $anv_{AV} = (x_{AV}, y_{AV}, z_{AV})$, resulting in 10 parameters for the AV at each time step.

Similarly, for the two NPC vehicles in the test environment, the same 10 basic parameters are observed for each. Considering the interaction between the AV and NPC vehicles, five additional relative parameters are defined for each NPC vehicle: $<dis_{rel}, yaw_{rel}, vel_{rel}, acc_{rel}, anv_{rel}>$. Among them, $dis_{rel}$, $vel_{rel}$, $acc_{rel}$, and $anv_{rel}$ are computed as the Euclidean norms between the corresponding parameters in the 10 basic parameter sets of the AV and the NPC vehicle, whereas $yaw_{rel}$ represents the difference in their heading angles, normalized to the range $[-180^\circ, 180^\circ]$.

Therefore, in a test environment, the observed state at each time step consists of 10 parameters for the AV and 15 parameters for each NPC vehicle (10 basic and 5 relative), resulting in a total of $10 + 15 \times 2 = 40$ parameters.

\textbf{Action Space.} We define each action as the configuration of an NPC vehicle’s driving behavior in the environment, to continuously simulate each NPC vehicle's driving process consistent with real-world behavior. As described in Section~\ref{subsubsec: parameters}, each NPC vehicle has 6 possible driving behaviors. Therefore, the action space consists of $6 \times 2 = 12$ possible actions, where each action corresponds to one potential driving behavior of a specific NPC vehicle.

\textbf{Vector Reward Function.} As mentioned in Section~\ref{subsubsec: requirements_objective_functions}, we focus on five key safety and functional requirements. Accordingly, the objective functions \dis, \ttc, \rc, \jerk, and \sd are defined to quantify their respective violation degrees or probabilities, where either lower or higher values may indicate more severe requirement violations depending on the specific metric. However, in \morl, higher rewards motivate the EQL agent to reproduce behaviors that yield those rewards. To drive the agent toward behaviors that are more likely to violate the requirements during testing, we therefore define the corresponding reward functions according to Equation~\ref{eq:reward_function}, ensuring that larger reward values consistently represent higher violation degrees.

$Reward_{Dis}$ denotes the reward function derived from the \dis objective function (Equation~\ref{eq:dis}), defined as follows:
\begin{equation}
Reward_{Dis} = 
\left\{
\begin{aligned}
    &10, && \text{if collision occurs} \\
    &1 - nor(\log(\min(Dis) + 1)), && \text{otherwise}
\end{aligned}
\right.
\end{equation}
where \dis denotes the minimum distance to surrounding vehicles computed at each instant, while $\min(Dis)$ represents the smallest of these distances observed within the entire time step. A logarithmic transformation is applied based on our pilot study, which showed that it better captures the variation in short-distance interactions.
We consider all requirements equally important, and thus apply a normalization function $nor(\cdot)$ to scale each reward into the range $[0, 1]$, ensuring comparability and fairness across objectives, which is reflected in this formulation and consistently applied to the other reward functions as well.
Besides, if a collision occurs, we set $Reward_{Dis}$ to 10 to strongly emphasize the critical severity of collisions, ensuring that the agent clearly distinguishes them from less severe interactions.

$Reward_{TTC}$ is defined based on the \ttc objective function (Equation~\ref{eq:ttc}), with its formulation given as follows:
\begin{equation}
Reward_{TTC} = 
\left\{
\begin{aligned}
    &10, && \text{if collision occurs} \\
    &1 - nor(\log(\min(TTC) + 1)), && \text{otherwise}
\end{aligned}
\right.
\end{equation}
where \ttc represents the minimum time to collision with surrounding vehicles at each instant, and $\min(TTC)$ denotes the minimum \ttc observed within the entire time step. As with $Reward_{Dis}$, based on our pilot study, we apply a logarithmic function to amplify smaller \ttc values, highlighting their relative impact. In addition, we use the $nor(\cdot)$ function to scale $Reward_{TTC}$ within the range $[0, 1]$, except in the case of a collision, where $Reward_{TTC}$ is set to 10.

The reward function $Reward_{RC}$ is constructed based on the \rc objective function (Equation~\ref{eq:rc}), which evaluates the route completion performance of the AV. Its formulation is given as follows:
\begin{equation}
Reward_{RC} = 
\left\{
\begin{aligned}
    &1 - nor(\Delta RC), && \text{if } \Delta RC \neq 0 \\
    &0, && \text{otherwise}
\end{aligned}
\right.
\end{equation}
where $\Delta RC$ represents the absolute difference in the AV’s \rc between the beginning and the end of the current time step.
We also apply the normalization function $nor(\cdot)$ to fairly constrain the values of $Reward_{RC}$ within the range $[0, 1]$.
Additionally, to prevent the generated scenario from causing the AV to remain stationary, leading to an invalid test, we set $Reward_{RC}$ to 0 when $\Delta RC$ remains consistently 0.

For the \jerk objective function (Equation~\ref{eq:jerk}), the corresponding reward function $Reward_{Jerk}$ is formulated to evaluate the ride comfort of passengers, as described below:
\begin{equation}
Reward_{Jerk} = nor(\max(Jerk))
\end{equation}
where \jerk represents the instantaneous rate of change of acceleration, reflecting the intensity of motion variations experienced by passengers, while $\max(Jerk)$ denotes the maximum \jerk observed within the current time step. The normalization function $nor(\cdot)$ is applied to scale $Reward_{Jerk}$ within the range $[0, 1]$.

To capture the speed difference with surrounding traffic, the reward function $Reward_{SD}$ is designed based on the \sd objective function (Equation~\ref{eq:sd}), as described below:
\begin{equation}
Reward_{SD} = nor(SD)
\end{equation}
where $nor(\cdot)$ serves as a normalization function that scales $SD$ into the range $[0, 1]$, ensuring consistency and comparability with other reward functions.
Note that the $SD$ in Equation~\ref{eq:sd} is computed for a single instant, whereas in the implementation of $Reward_{SD}$, it is derived from the speed data over the entire time step. Specifically, the speeds of all vehicles are first collected at each instant, averaged over the entire time step, and then compared to obtain the $SD$ value used in $Reward_{SD}$.

Based on the above definitions of individual reward functions, the overall vector reward function of \morl is constructed by selecting a subset of these component rewards according to the testing objective and scenario requirements.
Formally, let $\mathcal{S}\subseteq\{Dis,\,TTC,\,RC,\,Jerk,\,SD\}$ denote the selected set of objectives; the vector reward is then defined as the ordered tuple of the corresponding reward components:
\begin{equation}
\mathbf{R}_{\mathcal{S}} = \big[\, Reward_{o} \mid o \in \mathcal{S}\,\big].
\end{equation}
For example, if the test focuses on collision-related behaviors and ride comfort, we may select $\mathcal{S}=\{TTC,\,Jerk\}$ and set
\begin{equation}
\mathbf{R}_{\mathcal{S}} = [\, Reward_{TTC},\, Reward_{Jerk}\,],
\end{equation}
which guides the EQL agent to prioritize actions that influence time to collision and passenger comfort.

\subsubsection{Network Architecture, Parameter Settings, and Implementation}
\noindent To ensure the effectiveness of \morl and \sorl for a fair comparison, a suitable network architecture and corresponding parameter settings need to be selected. For this purpose, we employed an automated hyperparameter optimization framework, Optuna~\cite{akiba2019optuna}, and conducted a pilot study to optimize both the network architecture and parameter settings. The detailed implementation is available in our publicly available repository~\cite{MORLrepo}.

Based on the original EQL implementation~\cite{yang2019generalized} and a tool called \morl-Baselines~\cite{felten2024toolkit}, we implemented \morl. We adopt the $\epsilon
$-greedy policy~\cite{stadie2015incentivizing} to balance exploration and exploitation of the action space and utilize the prioritized experience replay~\cite{schaul2015prioritized} to store transition tuples. Additionally, we apply homotopy optimization~\cite{watson1989modern} for calculating the loss function. Meanwhile, \sorl employs the same implementation as DeepCollision, proposed by Lu et al.~\cite{lu2022learning}, which has demonstrated effectiveness in testing AVs using RL. 

To ensure comprehensive evaluation across different testing focuses, we designed a set of 10 pairwise objective combinations based on the reward definitions in Section~\ref{subsubsec:momdp}. Each combination includes two objectives (e.g., $\{TTC, Jerk\}$ for safety and comfort, or $\{Dis, RC\}$ for distance and efficiency), enabling us to investigate the interactions and trade-offs between different requirements while keeping the experimental complexity manageable.

For model training, based on Optuna's tuning results and our pilot study, we executed 1200 episodes, each consisting of 5 time steps, with 40 CARLA simulator minimum unit times allocated for each time step. 
To account for the stochastic nature of \morl and \sorl, during the model evaluation phase, we executed 100 episodes for both algorithms, freezing the training-related parameters while keeping the remaining parameters unchanged.

\subsubsection{Roads}\label{subsubsec:roads}
\noindent 
\morl and \sorl generate critical scenarios by configuring the driving environment of the AV throughout the driving task. In this work, we specify the driving tasks using different roads with various road structures. Each road has a starting and end point specifying the route of the AV, as well as the initial positions and driving directions of all NPC vehicles. Specifically, as Figure~\ref{fig:road} shows, we select six roads for the experiment, covering the six scenario categories on the CARLA Leaderboard~\cite{carlaleaderboard}. These categories are instances of the pre-crash scenarios from NHTSA pre-crash typology~\cite{najm2007pre}. 

\begin{figure*}[]\captionsetup[subfigure]{font=large}
	\centering
    \input{images/scenarios_n}
	\caption{Driving roads for specifying driving tasks. The gray arrows indicate the driving direction of the vehicles. The red solid dots represent the start or end points of the AV route and the blue arrows show the driving path of the AV.}
    \label{fig:road}
    \vspace{-25pt}
\end{figure*}

\textit{Road1} depicts a highway where the AV needs to drive from the leftmost lane to the rightmost lane, and then maintain the lane until reaching the end. The two NPC vehicles are initially positioned in the two rightmost lanes, one ahead of and one behind the AV, traveling in the same direction as the AV.

\textit{Road2} is a dual-way road with two opposite lanes, where the AV is expected to maintain its current lane until reaching the end. One NPC vehicle is initially positioned ahead of the AV, traveling in the same lane and direction as the AV, while the other NPC vehicle is generated in the opposite lane ahead of the AV, traveling in the opposite direction.

\textit{Road3} depicts a two-lane road where both lanes run in the same direction. The AV needs to merge from the right lane into the traffic flow in the left lane and continue along the left lane until reaching its destination. Both NPC vehicles are generated in the left lane, positioned one ahead of and one behind the AV, with all vehicles traveling in the same direction.

\textit{Road4} is a four-lane bidirectional road with two lanes in each direction. The AV needs to maintain its lane until reaching the end. All vehicles are generated on the same side of the road, traveling in the same direction. The AV is in the inner lane, with both NPC vehicles ahead. One is in the outer lane, and the other is in the inner lane, farther from the AV.

\textit{Road5} depicts a four-way intersection, with four lanes in each direction. The AV travels straight from the bottom lane through the intersection to the top lane. The two NPC vehicles cross the intersection from left to right and from right to left, respectively. All vehicles travel in the inner lanes.

\textit{Road6} is a two-lane one-way road, with the left lane severely damaged and in poor condition. The AV is in the left lane and must maintain its lane until reaching the end. While driving, the AV might lose control due to bad conditions and must recover to its original lane. One NPC vehicle is generated ahead of the AV in the same lane, and another NPC vehicle is generated behind the AV in the other lane.

\subsection{Evaluation Metrics}\label{subsec:metrics}
\noindent To answer the RQs, we define effectiveness metrics and diversity metrics to evaluate the performance of generated test scenarios. Effectiveness metrics quantify how well a scenario exposes potential risks or requirement violations in the ADS under test, while diversity metrics capture the variability among generated scenarios.

\subsubsection{Effectiveness Metrics}
Let $N$ be the total number of episodes and let the $e_{th}$ episode contain $T_e$ time steps. Denote by $x_{e,t}$ the value of a specific metric $x \in \mathcal{X}$ at time step $t$ in episode $e$, where 
\begin{equation}
\mathcal{X} = \{\min(Dis), \, \min(TTC), \, \Delta RC, \, \max(Jerk), \, SD\}.
\end{equation}
We compute each episode-level metric using an aggregation function $f(\cdot)$ defined as:
\begin{equation}
f\big(\{x_{e,t}\}_{t=1}^{T_e}\big) =
\begin{cases}
\sum_{t=1}^{T_e} x_{e,t}, & \text{if $x$ is } \Delta RC,\\[2pt]
\dfrac{1}{T_e}\sum_{t=1}^{T_e} x_{e,t}, & \text{otherwise}.
\end{cases}
\end{equation}
We then define the following metrics to evaluate scenario effectiveness:
\begin{itemize}[left=0pt,label=\textbullet]
    \item \textbf{Objective Value} ($\bm{OV}$) assesses the aggregated values of a single metric $x$ over all episodes, reflecting overall performance with respect to requirement violations and risk exposure.
    \item \textbf{Single-requirement Violation Count} ($\bm{\#SV}$) assesses the number of episodes violating the requirement associated with metric $x$, determined by comparing each episode-level value with its corresponding threshold, where the thresholds are set as $\theta_{Dis} = 5\,\text{m}$~\cite{wang2025autonomous,ro2020new}, $\theta_{TTC} = 1.0\,\text{s}$~\cite{lin2023commonroad}, $\theta_{RC} = 100\%$, $\theta_{Jerk} = 0.9\,\text{m/s}^3$~\cite{bae2019toward}, and $\theta_{SD} = 0.0\,\text{m/s}$.
    \item \textbf{Single-requirement Violation Severity} ($\bm{SVS}$) assesses the aggregated metric values computed only over the episodes that violate the corresponding requirement, indicating the single-requirement violation severity.
    \item \textbf{Multi-requirement Violation Count} ($\bm{\#MV}$) assesses the number of episodes simultaneously violating multiple requirements, used to capture multi-objective risk interactions.
    \item \textbf{Multi-requirement Violation Severity} ($\bm{MVS}$) assesses the aggregated metric values computed only over the episodes that simultaneously violate all corresponding requirements in the set $X \subseteq \mathcal{X}$, indicating the severity of multi-objective violations.
    \item \textbf{Collision Count} ($\bm{\#C}$) assesses the number of episodes in which a collision occurs, representing the most direct aspect of ADS failure.
\end{itemize}
Note that, in the subsequent sections, to simplify notation, each metric $x$ will be denoted using the name of its corresponding objective function. For example, if $x$ is $\min(Dis)$, it is simplified and denoted as $Dis$.

\subsubsection{Diversity Metrics}\label{subsubsec:div_metrics}
To evaluate the diversity of scenarios generated by the scenario generation algorithms, we introduce two complementary types of metrics: the \textit{behavior-level metric}, which quantifies variation in the sequences of actions (behaviors) chosen by the algorithm, and the \textit{scenario-level metric}, which quantifies variation in the physical and dynamic properties of the scenarios (e.g., positions, velocities, and other state variables).

The \textit{behavior-level metric} focuses on the diversity of the agent's action choices across all episodes. Specifically, a behavior denotes a single action taken by the agent, and a behavior sequence represents the set of actions executed in one episode. To capture different aspects of behavioral diversity, we introduce three metrics, defined as follows:
\begin{itemize}[left=0pt,label=\textbullet]
    \item \textbf{Unique Behavior Count} ($\bm{\#UB}$) assesses the number of unique behavior sequences across all episodes, capturing the variety of behavioral strategies employed by the agent.
    \item \textbf{Unique Behavior Diversity} ($\bm{UBD}$) assesses the average pairwise Hamming distance~\cite{hamming1950error} among only the unique behavior sequences across episodes, ignoring frequency, capturing the inherent diversity of distinct behaviors.
    \item \textbf{Weighted Behavior Diversity} ($\bm{WBD}$) assesses the average pairwise Hamming distance between all behavior sequences across episodes, weighted by their occurrence frequency, capturing the diversity of behaviors actually executed.
\end{itemize}

The \textit{scenario-level metric} is computed based on physical and dynamic parameters obtained from scenario executions, i.e., positions, rotations, velocities, accelerations, and angular velocities. These metrics characterize variations in the underlying scenario configurations and dynamics, providing a complementary physical perspective on diversity beyond the agent’s behavioral patterns.
We refer to this \textit{scenario-level metric} as \textbf{Scenario Diversity} ($\bm{SCD}$), representing the diversity across generated scenarios.

To compute \scd, each episode is represented as a time sequence. At every minimal time unit, we compute five types of pairwise physical parameter differences for all object pairs: relative position, relative rotation, relative velocity, relative acceleration, and relative angular velocity. With $n$ objects and $t$ time units, a single episode can therefore be expressed as a two-dimensional feature matrix of size $\left[\, n(n-1)/2 \times 5,\ t \,\right]$, capturing the evolving interaction relationships among objects within the generated scenario.
Subsequently, we compute the pairwise time-sequence distance between episodes using multi-dimensional Dynamic Time Warping (DTW)~\cite{berndt1994using}, thereby quantifying differences in their interaction patterns. DTW is a time-sequence similarity measure that allows nonlinear alignment along the temporal axis, making it suitable for comparing behavior sequences that may differ in pace or length.
Finally, we take the average DTW distance across all episode pairs as an indicator of the overall diversity of the generated scenarios, i.e., the \textit{scenario-level metric} \scd.

\subsection{Statistical Tests}\label{subsec:statistical}
\noindent 
We followed the guidelines~\cite{arcuri2011practical} to statistically analyze the results of the 100 evaluated episodes. 
For continuous numerical results, we applied the Mann-Whitney U test~\cite{wilcoxon1970critical} and the Vargha and Delaney \Atwelve statistics~\cite{vargha2000critique}. The Mann-Whitney U test assesses whether there is a significant difference between the distributions of the two independent samples, with a significance level set at 0.05 (indicating significance when $p < 0.05$). The Vargha and Delaney \Atwelve statistics is a non-parametric effect size that indicates the probability that the data in one group is greater than that in the other group. \Atwelve value greater than 0.5 suggests a high probability that the first group's data exceeds that of the second, while a value less than 0.5 indicates the opposite. If \Atwelve equals 0.5, it suggests no difference between the two groups. In accordance with the guideline~\cite{kitchenham2017robust}, 
we divide the magnitude of \Atwelve into four levels: \textit{negligible} (\Atwelve $\in$ (0.444, 0.556)), \textit{small} (\Atwelve $\in$ [0.556, 0.638) or (0.362, 0.444]), \textit{medium} (\Atwelve $\in$ [0.638, 0.714) or (0.286, 0.362]), \textit{large} (\Atwelve $\in$ [0.714, 1.0] or [0, 0.286]).
Given that some results of our experiment are dichotomous, e.g., a scenario violated both requirements (Yes or No), we employed the Fisher's exact test~\cite{fisher1970statistical} to determine the statistical significance of results based on the recommendations from~\cite{arcuri2011practical}, with a significance level of $p<0.05$. 
For the effect size, we chose odds ratios (OR)~\cite{szumilas2010explaining} based on an existing guide~\cite{arcuri2011practical}. When OR is 1, the two compared methods are equal. An OR greater than 1 suggests that the first method is more likely to be better than the second, while an OR less than 1 indicates the opposite. The further the OR deviates from 1, the stronger the effect.
According to the guideline~\cite{chen2010big}, ORs of 1.68, 3.47, and 6.71 correspond to small, medium, and large effects, respectively. Values of OR below 1 can be interpreted symmetrically on the logarithmic scale: $1/1.68\approx0.60$, $1/3.47\approx0.29$, and $1/6.71\approx0.15$, representing small, medium, and large effects in the opposite direction.

To investigate the impact of different objective combinations on scenario performance, we first assessed the correlations between each pair of objectives using Spearman's rank correlation coefficient~\cite{spearman1961proof} to support the analysis of their combined effects. The Spearman's rank correlation coefficient provides both a $p$-value, which indicates the significance of the correlation, and a correlation coefficient $\rho$, which reflects the direction and strength of the correlation. We set the significance level at 0.05, considering two variables to be significantly monotonically correlated only if the $p$-value is less than 0.05. The correlation coefficient $\rho$ ranges from -1 to +1, where $\rho > 0$ indicates a positive correlation, i.e., as one variable increases, the other tends to increase, and $\rho < 0$ suggests a negative correlation, i.e., as one variable increases, the other tends to decrease. According to the~\cite{schober2018correlation,zou2003correlation} guidelines, we consider $|\rho| \ge 0.5$ to indicate at least a moderate strength of monotonic correlation between the two variables.

Additionally, to mitigate the aggregated error probabilities arising from multiple comparisons, we applied the Holm–Bonferroni method~\cite{holm1979simple} to adjust the overall significance level, thereby controlling the family-wise error rate to a 5\% ($\alpha = 0.05$) level.

\section{Analysis of the Results and Discussion}\label{sec:results}

\subsection{Results for \textit{RQ1} -- Overall Performance Comparison}\label{subsec:RQ1}
\begin{figure}[]
    \centering
    \includegraphics[width=\linewidth]{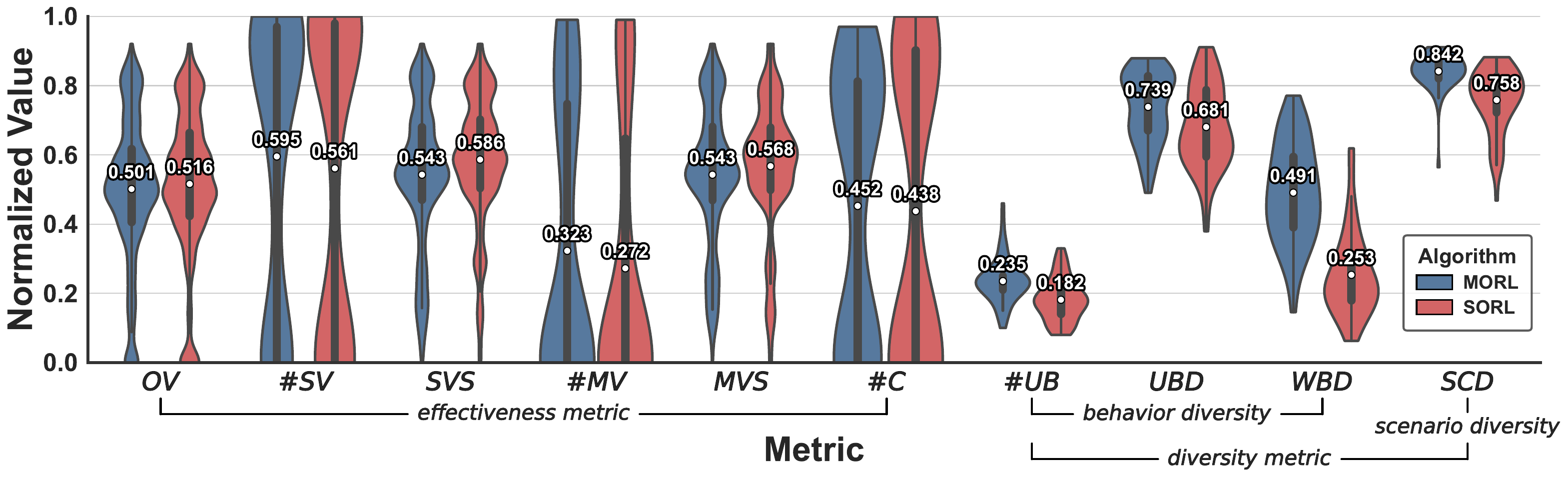}
    \caption{Normalized distribution results between \morl and \sorl over all objective combinations and road structures across effectiveness and diversity metrics. The mean represents the central tendency. \ov: Objective Value; \sv: Single-requirement Violation Count; \svs: Single-requirement Violation Severity; \mv: Multi-requirement Violation Count; \mvs: Multi-requirement Violation Severity; \cc: Collision Count; \ub: Unique Behavior Count; \ubd: Unique Behavior Diversity; \wbd: Weighted Behavior Diversity; \scd: Scenario Diversity. -- \textit{RQ1}}
    \label{fig:RQ1}
\end{figure}

\noindent
As introduced in Section~\ref{subsec:RQ}, \textit{RQ1} aims to study the overall performance differences between \sorl and \morl in generating critical scenarios across all the objective combinations and road structures. To answer this RQ, we evaluate the overall performance of \morl and \sorl using the effectiveness and diversity metrics defined in Section~\ref{subsec:metrics}, with summarized distribution results shown in Figure~\ref{fig:RQ1}. In addition, as described in Section~\ref{subsec:statistical}, we conduct statistical significance analysis on the results, which are presented in Table~\ref{tab:tabStatisticRQ1}.

\subsubsection{Descriptive Analysis}
Figure~\ref{fig:RQ1} illustrates the overall distribution of effectiveness and diversity metrics for \morl and \sorl across all 10 objective combinations and 6 road structures. Results are normalized to [0,1] for comparability, using the same calculation as the reward functions defined for each objective in Section~\ref{subsubsec:momdp}. Higher values indicate better performance.

\vspace{3pt}
\noindent \textit{\textbf{Analysis Regarding Effectiveness Metrics}}.
For the effectiveness metrics shown in Figure~\ref{fig:RQ1}, i.e., \ov, \sv, \svs, \mv, \mvs, and \cc, the distributions of \morl and \sorl are largely similar, indicating comparable overall effectiveness between the two approaches, while some differences can be observed in their overall average values. 
Specifically, for the mean \textit{Objective Value} (\ov), \sorl achieves a slightly higher \ov than \morl (0.516 vs. 0.501), indicating that \sorl tends to achieve higher effectiveness on average across all generated scenarios under different objective combinations and road structures, although the performance difference remains relatively small (i.e., 0.015).

Moreover, we further analyze the performance differences between \morl and \sorl from the perspective of requirement violations. For single-requirement violations, \morl obtains a higher average \textit{Single-Requirement Violation Count} (\sv) than \sorl (0.595 vs. 0.561), indicating that \morl generates more scenarios violating a specific requirement. However, \sorl achieves a higher mean \textit{Single-Requirement Violation Severity} (\svs) than \morl (0.586 vs. 0.543), suggesting that the violation scenarios generated by \sorl tend to exhibit greater severity.
For multi-requirement violations, \morl shows a higher mean \textit{Multi-Requirement Violation Count} (\mv) than \sorl (0.323 vs. 0.272), while \sorl demonstrates a higher average \textit{Multi-Requirement Violation Severity} (\mvs) than \morl (0.568 vs. 0.543). This suggests that \morl tends to generate more scenarios simultaneously violating multiple interdependent requirements, whereas \sorl tends to produce violations of greater severity.
Therefore, based on the above findings, we conclude that \morl tends to generate a larger number of requirement-violation test scenarios, whereas the violation scenarios generated by \sorl generally reflect more severe violations.

Finally, we report the \textit{Collision Count} (\cc) for both \morl and \sorl. Although this metric is directly relevant to safety objectives, we compute it for all objective combinations to maintain consistency in evaluation.
As shown in Figure~\ref{fig:RQ1}, the mean values of \morl (0.452) and \sorl (0.438) on \cc are very close, with \morl only slightly higher than \sorl (0.014). This suggests that both methods are capable of effectively generating collision scenarios, while neither method demonstrates a clear overall advantage on this metric.

\vspace{3pt}
\noindent \textit{\textbf{Analysis Regarding Diversity Metrics}}.
In Figure~\ref{fig:RQ1}, it is evident that \morl tends to produce higher values across both \textit{behavior-level diversity} (\ub, \ubd, and \wbd) and \textit{scenario-level diversity} (\sd), although its distributions partially overlap with those of \sorl. Since higher values indicate better performance, the distributions suggest that \morl generates more diverse test scenarios overall. This observation is further supported by the average values of all diversity metrics, which are consistently higher for \morl than for \sorl.

\subsubsection{Statistical Analysis}

To further analyze the statistical significance of the effectiveness and diversity results, we conduct statistical tests following the design described in Section~\ref{subsec:statistical}. Specifically, for \ov, \svs, \mvs, \ubd, \wbd, and \scd, we apply the Mann-Whitney U test with the Vargha and Delaney \Atwelve effect size. For \sv, \mv, \cc, and \ub, we use Fisher's exact test together with the OR effect size. The statistical test results are summarized in Table~\ref{tab:tabStatisticRQ1}.

\begin{table}[]
\centering
\caption{Overall statistical results between \morl and \sorl for all objective combinations and road structures across effectiveness and diversity metrics. \ov: Objective Value; \sv: Single-requirement Violation Count; \svs: Single-requirement Violation Severity; \mv: Multi-requirement Violation Count; \mvs: Multi-requirement Violation Severity; \cc: Collision Count; \ub: Unique Behavior Count; \ubd: Unique Behavior Diversity; \wbd: Weighted Behavior Diversity; \scd: Scenario Diversity. $e$ represents the effect size. \underline{Underlined} denotes non-significant differences. \textbf{Bold} indicates a moderate or larger effect size. -- \textit{RQ1}}
\label{tab:tabStatisticRQ1}
\vspace{10pt}
\resizebox{0.8\linewidth}{!}{
\begin{tabular}{lllllllllll}
\toprule
\multirow{2.5}{*}{\textbf{\begin{tabular}[c]{@{}l@{}}Statistical\\ Measure\end{tabular}}} & \multicolumn{10}{l}{\textbf{Metric}} \\ \cmidrule(r){2-11}
 & $\bm{OV}$ & $\bm{\#SV}$ & $\bm{SVS}$ & $\bm{\#MV}$ & $\bm{MVS}$ & $\bm{\#C}$ & $\bm{\#UB}$ & $\bm{UBD}$ & $\bm{WBD}$ & $\bm{SCD}$ \\ \cmidrule(r){1-11}
$\bm{p}$ & \textless{}0.01 & \textless{}0.01 & \textless{}0.01 & \textless{}0.01 & \textless{}0.01 & \underline{$\ge$0.05} & \textless{}0.01 & \textbf{\textless{}0.01} & \textbf{\textless{}0.01} & \textbf{\textless{}0.01} \\
$\bm{e}$ & 0.476 & 1.151 & 0.434 & 1.274 & 0.455 & \underline{1.063} & 1.388 & \textbf{0.649} & \textbf{0.895} & \textbf{0.812} \\ \bottomrule
\end{tabular}
}
\vspace{-5pt}
\end{table}

\vspace{3pt}
\noindent \textit{\textbf{Statistical Results for Effectiveness Metrics}}.
Regarding \ov, \svs, and \mvs, Table~\ref{tab:tabStatisticRQ1} reveals statistically significant differences between \morl and \sorl ($p<0.05$). The corresponding effect sizes (\Atwelve) are all below 0.5, indicating that \sorl achieves higher values on these metrics. However, the effect sizes are all negligible to small (\Atwelve $\in$ (0.362, 0.556)), meaning that the two approaches perform comparably in practice.
This observation is consistent with the distribution results shown in Figure~\ref{fig:RQ1}, where \sorl achieves slightly higher mean values than \morl on these metrics, while the gaps remain small. 
This suggests that \morl and \sorl demonstrate comparable performance in terms of objective values and violation severity, with \sorl showing a slight tendency to generate more severe violations.

For \sv and \mv, the statistical tests indicate that \morl causes significantly more single/multi-requirement violations than \sorl, with effect sizes ($OR$) greater than 1 but below 3.47 (medium level), suggesting a statistically significant yet practically small advantage of \morl over \sorl.
This observation is consistent with the results shown in Figure~\ref{fig:RQ1}, where the two approaches exhibit similar performance on \sv and \mv, with \morl being only marginally higher than \sorl. 
Overall, these results suggest that \morl and \sorl exhibit generally similar performance regarding the number of generated single- or multi-requirement violation scenarios, and \morl tends to reveal slightly more such scenarios.

For \cc, the statistical result shows no statistically significant difference between \morl and \sorl, meaning that the two approaches demonstrate comparable capabilities in triggering collisions. This finding is also consistent with the observation from Figure~\ref{fig:RQ1}.

\vspace{3pt}
\noindent \textit{\textbf{Statistical Results for Diversity Metrics}}.
For the diversity metrics, Table~\ref{tab:tabStatisticRQ1} shows that \morl significantly outperforms \sorl on \ubd, \wbd, and \scd, while for \ub, \morl performs better than \sorl, although the difference is statistically significant, the effect size is negligible. These results indicate that, overall, \morl generates test scenarios with significantly higher diversity than \sorl.
To understand why the observed difference for \ub is statistically significant but practically small, we further analyze how each diversity metric is calculated.
Recall that in Section~\ref{subsubsec:div_metrics}, \ub counts the number of unique behavior sequences of NPC vehicles across all generated scenarios and therefore evaluates diversity at the sequence level. \ubd and \wbd further consider the differences between individual actions within behavior sequences, while \scd captures scenario-level diversity by measuring the dynamic differences in NPC interactions based on the DTW distance of their physical feature time series during scenario execution.
As a result, these metrics differ in their sensitivity to variations in behavior sequences and interaction dynamics.
Therefore, it is possible that \morl and \sorl achieve similar results on \ub, while showing significant differences on \ubd, \wbd, and \scd. This suggests that although \sorl may generate a similar number of unique NPC behavior sequences as \morl, these sequences tend to be more similar in terms of their action patterns and interaction dynamics, leading to lower diversity values on \ubd, \wbd, and \scd.

\vspace{-5pt}
\begin{center}
\fcolorbox{black}{gray!10}{
\parbox{0.97\columnwidth}{
\textbf{Conclusion for \textit{RQ1}}:
Overall, \morl and \sorl demonstrate comparable performance regarding effectiveness metrics, with a slight difference observed. Specifically, \morl tends to generate more requirement-violation scenarios, whereas the violations produced by \sorl tend to be more severe. Both approaches exhibit similar capability in generating collision scenarios, with no clear advantage for either approach. In terms of diversity, \morl generates test scenarios with significantly higher overall diversity than \sorl, both at the behavior and scenario levels.
}}
\end{center}

\subsection{Results for \textit{RQ2} -- Impact of Objective Combinations}\label{subsec:RQ2}
\begin{figure}[]
    \centering
    \includegraphics[width=0.85\linewidth]{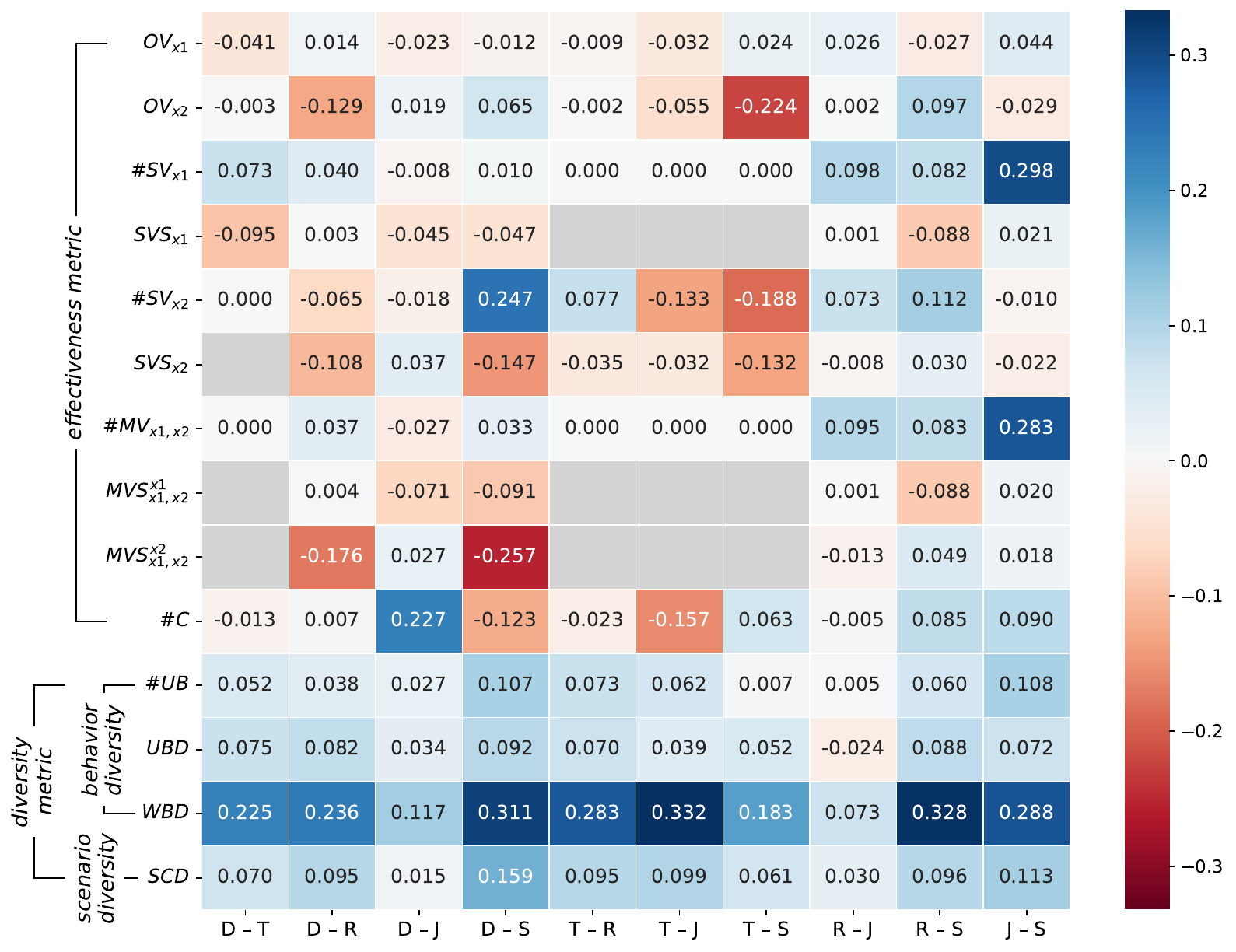}
    \caption{Heatmap of normalized average performance differences between \morl and \sorl across metrics and objective combinations. \ov: Objective Value; \sv: Single-requirement Violation Count; \svs: Single-requirement Violation Severity; \mv: Multi-requirement Violation Count; \mvs: Multi-requirement Violation Severity; \cc: Collision Count; \ub: Unique Behavior Count; \ubd: Unique Behavior Diversity; \wbd: Weighted Behavior Diversity; \scd: Scenario Diversity. ``$x1$--$x2$" represents a pairwise objective combination, where ``$x1$" denotes one objective and ``$x2$" denotes another. D: \dis, T: \ttc, R: \rc, J: \jerk, and S: \sd. Blue indicates \morl outperforms \sorl; Red indicates \sorl performs better. Gray cells denote invalid data. -- \textit{RQ2}}
    \label{fig:RQ2}
    \vspace{-5pt}
\end{figure}

\noindent
\textit{RQ2} investigates the impact of different objective combinations on the comparative performance of \sorl and \morl in test scenario generation.
To answer this RQ, we construct heatmaps to visualize the performance differences between \morl and \sorl across all effectiveness and diversity metrics under each objective combination (Figure~\ref{fig:RQ2}), and count the number of effectiveness metrics in which one method outperforms the other (Table~\ref{tab:tabCountRQ2}). Statistical significance tests are conducted following Section~\ref{subsec:statistical}, with results reported in Table~\ref{tab:tabStatisticRQ2}. In addition, we analyze the correlation between pairwise objectives under different combinations, as summarized in Table~\ref{tab:tabCorrelationSpearman}.

\subsubsection{Overall Analysis}\label{subsubsec:RQ2overall}
Figure~\ref{fig:RQ2} presents heatmaps illustrating the performance differences between \morl and \sorl across different effectiveness and diversity metrics under various objective combinations. Overall, \morl outperforms \sorl on all diversity metrics (i.e., \ub, \ubd, \wbd, and \scd), except for \ubd under the R--J objective combination, where the difference is small (-0.024). This suggests that \morl consistently generates more diverse test scenarios across objective combinations.
As most diversity metrics are computed over all generated scenarios, distinguishing objective combinations without considering road structures results in limited data per combination (fewer than 10 samples), making statistical testing less reliable. Thus, we do not perform statistical analysis for the diversity metrics.
For the remaining effectiveness metrics, no clear overall advantage is observed between the two methods from the heatmaps; therefore, further analysis under specific objective combinations is required.
Notably, zero or incomputable differences only occur in metrics related to requirement violations and all involve \ttc, suggesting that the evaluated ADS is relatively robust to \ttc. This may be due to the strict handling of the corresponding requirement by the evaluated ADS, which limits the optimization space of \ttc and makes such violations difficult to trigger.

\subsubsection{Objective Combination-level Analysis}
\begin{table}[]
\centering
\caption{Outperformance counts between \morl and \sorl across effectiveness metrics for each objective combination. ``$x1$--$x2$" represents a pairwise objective combination, where ``$x1$" denotes one objective and ``$x2$" denotes another. D: \dis, T: \ttc, R: \rc, J: \jerk, and S: \sd. \textbf{Bold} indicates the better result. -- \textit{RQ2}}
\label{tab:tabCountRQ2}
\vspace{10pt}
\resizebox{0.7\linewidth}{!}{
\begin{tabular}{lllllllllll}
\toprule
\multirow{2.4}{*}{\textbf{Algorithm}} & \multicolumn{10}{l}{\textbf{Objective Combination}} \\ \cmidrule(r){2-11}
 & \textbf{D--T} & \textbf{D--R} & \textbf{D--J} & \textbf{D--S} & \textbf{T--R} & \textbf{T--J} & \textbf{T--S} & \textbf{R--J} & \textbf{R--S} & \textbf{J--S} \\ \cmidrule(r){1-11}
\morl & 1 & \textbf{6} & 4 & 4 & 1 & 0 & 2 & \textbf{7} & \textbf{7} & \textbf{7} \\
\sorl & \textbf{4} & 4 & \textbf{6} & \textbf{6} & \textbf{4} & \textbf{5} & \textbf{3} & 3 & 3 & 3 \\ \bottomrule
\end{tabular}
}
\end{table}
For each objective combination, we count the number of effectiveness metrics on which one algorithm outperforms the other based on Figure~\ref{fig:RQ2} (outperformance counts), with the results shown in Table~\ref{tab:tabCountRQ2}. Overall, \sorl achieves more outperformance counts in 6 objective combinations, while \morl performs better in the remaining 4.
Specifically, when the objective combination includes \ttc, \sorl consistently outperforms \morl. Among the combinations that do not include \ttc, \morl performs better when \rc is present. Further excluding these combinations, \sorl shows better performance when \dis is included. For objective combinations involving \jerk or \sd, neither method exhibits a clear performance preference.
These results indicate that, across all objective combinations, different objectives exhibit different levels of priority in influencing algorithm performance. \ttc has the most pronounced impact, with \sorl showing superior performance in combinations that include this objective. This is followed by \rc, which also demonstrates a relatively strong influence, with \morl performing better when it is included. A similar pattern is observed for \dis, and it appears to be more favorable to \sorl. In contrast, for combinations involving \jerk or \sd, both methods show comparable performance.
These conclusions are drawn from the overall analysis across all effectiveness metrics for each objective combination. To better understand the underlying reasons, we perform a more detailed analysis of individual metrics as shown in the following section.

\subsubsection{Metric-level Analysis}
To further uncover the underlying reasons behind the impact of different objective combinations on the comparative performance of \morl and \sorl, we integrate the results from Figure~\ref{fig:RQ2}, Table~\ref{tab:tabStatisticRQ2}, and Table~\ref{tab:tabCorrelationSpearman}, and conduct a detailed analysis by examining individual effectiveness metrics.

\begin{table}[]
\centering
\caption{Statistical results comparing \morl and \sorl across pairwise objective combinations and effectiveness metrics. ``$x1$--$x2$" represents a pairwise objective combination, where ``$x1$" denotes one objective and ``$x2$" denotes another. D: \dis, T: \ttc, R: \rc, J: \jerk, and S: \sd. \ov: Objective Value; \sv: Single-requirement Violation Count; \svs: Single-requirement Violation Severity; \mv: Multi-requirement Violation Count; \mvs: Multi-requirement Violation Severity; \cc: Collision Count. $e$ represents the effect size. ``NA" means not available. \textbf{Bold} indicates a moderate or larger effect size. -- \textit{RQ2}}
\label{tab:tabStatisticRQ2}
\vspace{10pt}
\resizebox{0.7\linewidth}{!}{
\renewcommand{\arraystretch}{1.2}
\begin{tabular}{ccccccccccc}
\toprule
\multirow{2.5}{*}{\textbf{Metric}} & \multicolumn{2}{c}{\textbf{D--T}} & \multicolumn{2}{c}{\textbf{D--R}} & \multicolumn{2}{c}{\textbf{D--J}} & \multicolumn{2}{c}{\textbf{D--S}} & \multicolumn{2}{c}{\textbf{T--R}} \\ \cmidrule(r){2-11}
 & $\bm{p}$ & $\bm{e}$ & $\bm{p}$ & $\bm{e}$ & $\bm{p}$ & $\bm{e}$ & $\bm{p}$ & $\bm{e}$ & $\bm{p}$ & $\bm{e}$ \\ \cmidrule(r){1-11}
$\bm{OV_{x1}}$ & \textless{}0.01 & 0.552 & $\ge$0.05 & 0.472 & \textless{}0.05 & 0.534 & $\ge$0.05 & 0.514 & $\ge$0.05 & 0.493 \\
$\bm{OV_{x2}}$ & $\ge$0.05 & 0.514 & \textbf{\textless{}0.01} & \textbf{0.684} & \textless{}0.01 & 0.551 & \textless{}0.01 & 0.554 & $\ge$0.05 & 0.496 \\
$\bm{\#SV_{x1}}$ & \textless{}0.05 & 1.343 & $\ge$0.05 & 1.192 & $\ge$0.05 & 0.966 & $\ge$0.05 & 1.046 & $\ge$0.05 & NA \\
$\bm{SVS_{x1}}$ & \textbf{\textless{}0.01} & \textbf{0.771} & $\ge$0.05 & 0.510 & \textbf{\textless{}0.01} & \textbf{0.671} & \textbf{\textless{}0.01} & \textbf{0.656} & NA & NA \\
$\bm{\#SV_{x2}}$ & $\ge$0.05 & NA & \textless{}0.01 & 0.427 & $\ge$0.05 & 0.889 & \textless{}0.01 & 3.112 & \textless{}0.01 & 2.057 \\
$\bm{SVS_{x2}}$ & NA & NA & \textbf{\textless{}0.01} & \textbf{0.682} & \textless{}0.01 & 0.613 & \textbf{\textless{}0.01} & \textbf{0.335} & \textless{}0.05 & 0.545 \\
$\bm{\#MV_{x1,x2}}$ & $\ge$0.05 & NA & $\ge$0.05 & 1.175 & $\ge$0.05 & 0.882 & $\ge$0.05 & 1.246 & $\ge$0.05 & NA \\
$\bm{MVS_{x1,x2}^{x1}}$ & NA & NA & $\ge$0.05 & 0.505 & \textbf{\textless{}0.01} & \textbf{0.796} & \textbf{\textless{}0.01} & \textbf{0.745} & NA & NA \\
$\bm{MVS_{x1,x2}^{x2}}$ & NA & NA & \textbf{\textless{}0.01} & \textbf{0.933} & \textless{}0.01 & 0.600 & \textbf{\textless{}0.01} & \textbf{0.063} & NA & NA \\
$\bm{\#C}$ & $\ge$0.05 & 0.932 & $\ge$0.05 & 1.028 & \textless{}0.01 & 2.573 & \textless{}0.01 & 0.604 & $\ge$0.05 & 0.909 \\ \cmidrule(r){1-11}
& \multicolumn{2}{c}{\textbf{T--J}} & \multicolumn{2}{c}{\textbf{T--S}} & \multicolumn{2}{c}{\textbf{R--J}} & \multicolumn{2}{c}{\textbf{R--S}} & \multicolumn{2}{c}{\textbf{J--S}} \\ \cmidrule(r){1-11}
$\bm{OV_{x1}}$ & \textless{}0.01 & 0.562 & $\ge$0.05 & 0.481 & \textless{}0.05 & 0.462 & \textless{}0.05 & 0.540 & \textbf{\textless{}0.01} & \textbf{0.684} \\
$\bm{OV_{x2}}$ & \textbf{\textless{}0.01} & \textbf{0.333} & \textbf{\textless{}0.01} & \textbf{0.287} & $\ge$0.05 & 0.495 & \textless{}0.01 & 0.614 & $\ge$0.05 & 0.494 \\
$\bm{\#SV_{x1}}$ & $\ge$0.05 & NA & $\ge$0.05 & NA & \textless{}0.01 & 1.677 & \textless{}0.01 & 1.918 & \textbf{\textless{}0.01} & \textbf{3.499} \\
$\bm{SVS_{x1}}$ & NA & NA & NA & NA & $\ge$0.05 & 0.519 & \textless{}0.01 & 0.612 & \textbf{\textless{}0.01} & \textbf{0.682} \\
$\bm{\#SV_{x2}}$ & \textbf{\textless{}0.01} & \textbf{0.199} & \textbf{\textless{}0.01} & \textbf{0.066} & \textless{}0.01 & 2.125 & \textless{}0.01 & 2.760 & $\ge$0.05 & 0.495 \\
$\bm{SVS_{x2}}$ & \textless{}0.01 & 0.377 & \textbf{\textless{}0.01} & \textbf{0.348} & \textless{}0.01 & 0.447 & \textless{}0.01 & 0.577 & $\ge$0.05 & 0.499 \\
$\bm{\#MV_{x1,x2}}$ & $\ge$0.05 & NA & $\ge$0.05 & NA & \textless{}0.01 & 1.624 & \textless{}0.01 & 1.939 & \textless{}0.01 & 3.257 \\
$\bm{MVS_{x1,x2}^{x1}}$ & NA & NA & NA & NA & $\ge$0.05 & 0.519 & \textless{}0.01 & 0.612 & \textbf{\textless{}0.01} & \textbf{0.680} \\
$\bm{MVS_{x1,x2}^{x2}}$ & NA & NA & NA & NA & \textless{}0.01 & 0.409 & \textless{}0.01 & 0.595 & \textless{}0.01 & 0.576 \\
$\bm{\#C}$ & \textless{}0.01 & 0.499 & \textless{}0.05 & 1.308 & $\ge$0.05 & 0.497 & \textbf{\textless{}0.01} & \textbf{12.250} & \textbf{\textless{}0.01} & \textbf{12.978} \\ \bottomrule
\end{tabular}
}
\end{table}

\begin{table}[]
\centering
\caption{Spearman’s rank correlation coefficients among pairwise objective combinations. ``$x1$--$x2$" represents a pairwise objective combination, where ``$x1$" denotes one objective and ``$x2$" denotes another. D: \dis, T: \ttc, R: \rc, J: \jerk, and S: \sd. \textbf{Bold} indicates moderate or strong monotonic correlations. \underline{Underlined} denotes non-significant correlations. -- \textit{RQ2}}
\label{tab:tabCorrelationSpearman}
\vspace{10pt}
\resizebox{0.8\linewidth}{!}{
\begin{tabular}{lllllllllll}
\toprule
\multirow{2.4}{*}{\textbf{\begin{tabular}[c]{@{}l@{}}Statistical\\ Measure\end{tabular}}} & \multicolumn{10}{l}{\textbf{Objective Combination}} \\ \cmidrule(r){2-11}
 & \textbf{D--T} & \textbf{D--R} & \textbf{D--J} & \textbf{D--S} & \textbf{T--R} & \textbf{T--J} & \textbf{T--S} & \textbf{R--J} & \textbf{R--S} & \textbf{J--S} \\ \cmidrule(r){1-11}
$\bm{p}$ & \textless{}0.01 & \underline{$\ge$0.05} & \textless{}0.01 & \textless{}0.01 & \textless{}0.05 & \textless{}0.01 & \textless{}0.01 & \textless{}0.01 & \textbf{\textless{}0.01} & \textbf{\textless{}0.01} \\
\textit{$\bm{\rho}$} & -0.237 & \underline{0.038} & 0.327 & -0.103 & 0.076 & -0.156 & -0.128 & -0.480 & \textbf{-0.539} & \textbf{-0.527} \\ \bottomrule
\end{tabular}
}
\end{table}

\vspace{3pt}
\noindent \textit{\textbf{Analysis Regarding Objective Value Metric}}.
Regarding the \ov results shown in Figure~\ref{fig:RQ2}, \sorl outperforms \morl on both objectives in most \ttc-involved combinations, i.e., D--T, T--R, and T--J. For T--S, \sorl performs better on \sd but slightly worse on \ttc. The statistical results in Table~\ref{tab:tabStatisticRQ2} further indicate that no significant differences are observed on \ttc across all \ttc-involved combinations. In contrast, significant differences exist for the other objectives (except \rc), with \sorl demonstrating medium effect size advantages on \jerk and \sd.
These results suggest that, for objective combinations involving \ttc, the two methods achieve similar performance on \ttc (with \sorl being slightly better), while \sorl is more effective in optimizing the other objectives.
Furthermore, the correlation analysis in Table~\ref{tab:tabCorrelationSpearman} shows that \ttc is weakly correlated with the other objectives. Combined with the analysis in Section~\ref{subsubsec:RQ2overall}, this suggests that the selected ADS imposes strict control over \ttc, thereby limiting its optimization space and obscuring its interaction with other objectives (though \ttc may also be inherently weakly correlated with them).
Under this condition, both algorithms exhibit similar performance on \ttc when optimized jointly with another objective, while improvements mainly occur on the other objective.
Consequently, for \ttc-involved combinations, the optimization problem effectively approaches a single-objective setting. In such cases, \sorl is better suited as it was originally designed for single-objective optimization, while \morl, which emphasizes trade-offs, may be less effective.

For the remaining objective combinations, neither method consistently outperforms the other on both objectives in Figure~\ref{fig:RQ2}. However, Table~\ref{tab:tabStatisticRQ2} shows that significant differences between \sorl and \morl are observed for the \rc in D--R, the \jerk in T--J and J--S, and the \sd in T--S.

\begin{itemize}[left=0pt,label=\textbullet]
\item For the \rc objective, a statistically significant advantage of \sorl over \morl is observed only in the D--R combination, while no significant difference is found on \dis. In the other \rc-involved combinations, although some statistically significant differences are observed, their effect sizes are small, indicating negligible practical differences. Further examining Table~\ref{tab:tabCorrelationSpearman}, D--R is the only combination in which the two objectives show no significant correlation. This suggests that \rc and \dis are relatively independent in this setting, allowing each algorithm’s optimization preference on \rc to be more directly reflected. This may explain why a significant difference is observed for \rc in D--R, but not for \dis, with differences in other combinations remaining limited.

\item For the \jerk objective, \sorl significantly outperforms \morl in the T--J combination, whereas \morl shows a significant advantage in the J--S combination. In \ttc-involved combinations, the optimization of \ttc is constrained, making the other objective more prominent, which helps explain \sorl’s advantage on \jerk in T--J. In J--S, Table~\ref{tab:tabCorrelationSpearman} shows a significant negative correlation between \jerk and \sd, indicating a trade-off. The significant advantage of \morl on \jerk suggests that it prefers the trade-off direction associated with \jerk in this combination. The lack of significance on \sd indicates that the two algorithms perform comparably on this objective, or that the trade-off has a limited impact. It cannot be ruled out that \sorl handles multi-objective trade-offs less effectively than \morl in this combination, which may lead to the observed difference on \jerk.

\item For the \sd objective, \sorl shows significantly better performance than \morl in the T--S combination, which may be due to the presence of \ttc, causing \sorl to focus more effectively on \sd. In the other objective combinations, differences between the two algorithms are negligible or small. These results suggest that, when focusing on \sd in T--S, \sorl’s optimization is more effective, while in general, the two algorithms exhibit similar performance across most combinations.

\end{itemize}

\vspace{-13pt}
\noindent \textit{\textbf{Analysis Regarding Single-requirement Violation Metrics}}.
From the $\#SV_{x1}$ and $\#SV_{x2}$ results in Figure~\ref{fig:RQ2}, for \textbf{objective combinations involving \dis}, \morl generally achieves higher values on the $\#SV_{D}$ metric, indicating a greater ability to generate scenarios violating requirement \textit{R1}. The only exception is D--J, where \morl performs slightly worse, though the difference is small. However, Table~\ref{tab:tabStatisticRQ2} shows that no/negligible significant differences are observed across these combinations. This suggests that \morl and \sorl exhibit comparable performance on $\#SV_{D}$, with \morl showing a slight advantage.
Additionally, in the D--J combination, \sorl outperforms \morl on both \sv metrics, but the differences from Table~\ref{tab:tabStatisticRQ2} are not statistically significant, indicating comparable performance.
In the D--S combination, \morl achieves higher values on both \sv metrics. Statistical results show a significant advantage of \morl on $\#SV_{S}$ with a small effect size, while no significant difference is observed on $\#SV_{D}$. This means that the two methods perform similarly for the D--S combination, with \morl showing a slight advantage on \sd.
Considering the $SVS_{x1}$ and $SVS_{x2}$ results in Figure~\ref{fig:RQ2}, \sorl outperforms \morl in most \dis-involved combinations, with slight exceptions on \dis in D--R and \jerk in D--J. Table~\ref{tab:tabStatisticRQ2} further shows that these advantages are generally statistically significant, except for \dis in D--R, while the difference on \jerk in D--J has a small effect size. This suggests that \sorl tends to generate scenarios with more severe requirement violations than \morl in \dis-related combinations.

For \textbf{objective combinations involving \rc}, Figure~\ref{fig:RQ2} shows that \sorl achieves higher $\#SV_{R}$ values only in the D--R combination, whereas \morl attains higher values in the other combinations. When considering both objectives, \morl also achieves higher $\#SV$ values for the corresponding objectives in these combinations. Although these differences are statistically significant (Table~\ref{tab:tabStatisticRQ2}), the effect sizes are negligible or small, indicating overall comparable performance, with \morl tending to generate a slightly larger number of violating scenarios in most \rc-involved combinations.
Notably, Table~\ref{tab:tabCorrelationSpearman} shows that \dis and \rc are uncorrelated in the D--R combination, unlike all other objective pairs. This suggests that when objectives are independent, \sorl may be more effective, whereas \morl tends to perform better when objectives are correlated.
For all \rc-involved combinations, the $SVS_{x1}$ and $SVS_{x2}$ results in Figure~\ref{fig:RQ2} show that \sorl generally achieves better performance on \rc than \morl, except for a minor difference in R--J. However, neither method consistently outperforms the other on both objectives.
Statistical results further indicate that, except for a medium effect size observed on \rc in D--R, differences are either not significant or have small effect sizes. This suggests that both methods exhibit comparable performance in terms of violation severity for \rc-related combinations, with \sorl being slightly more effective on \rc.
For the D--R combination, consistent with previous analysis, the two objectives are uncorrelated, under which \sorl appears to be more effective, leading to its significant advantage on \rc.

For \textbf{objective combinations involving \jerk}, Figure~\ref{fig:RQ2} shows that \sorl achieves better performance on $\#SV_{J}$ in the D--J and T--J combinations, whereas \morl performs better in the R--J and J--S combinations. Statistical results in Table~\ref{tab:tabStatisticRQ2} indicate that most $\#SV_{J}$ differences are significant, but only the T--J and J--S combinations exhibit medium effect sizes, suggesting that meaningful differences mainly occur in these two cases.
Considering both objectives, \sorl outperforms \morl on all non-zero $\#SV$ metrics in D--J and T--J, whereas \morl performs better on both objectives in R--J. In J--S, \morl achieves better performance on $\#SV_{J}$, while \sorl performs better on $\#SV_{S}$. The statistical results further show that only the differences on \jerk in T--J and J--S have medium effect sizes, while the others remain negligible or small.
For the T--J combination, a significant difference is observed only on $\#SV_{J}$, which may be attributed to the restricted optimization space of \ttc, making the problem closer to a single-objective setting where \sorl performs better.
For the J--S combination, \morl shows a significant advantage on $\#SV_{J}$, while no significant difference is observed on $\#SV_{S}$. This is consistent with the previously identified trade-off between \jerk and \sd, where \morl tends to prefer the optimization direction associated with \jerk, while both methods perform comparably on \sd.
Overall, these results indicate that the number of violation scenarios generated for \jerk by \sorl and \morl is highly dependent on the specific objective combination, with meaningful differences primarily observed in T--J and J--S, where problem characteristics such as \ttc constraints and objective trade-offs play a key role.
Moreover, the $SVS_{x1}$ and $SVS_{x2}$ results in Figure~\ref{fig:RQ2} and Table~\ref{tab:tabStatisticRQ2} show different comparative patterns across combinations.
In D--J, \sorl performs better on \dis, while \morl performs better on \jerk; the difference is significant for \dis but small for \jerk.
In T--J, only \jerk is valid, where \sorl outperforms \morl, but the difference is not practically significant.
In R--J, \morl performs better on \rc and \sorl on \jerk; the difference is not significant for \rc, but significant with a small effect size for \jerk.
In J--S, \morl significantly outperforms \sorl on \jerk, while the difference on \sd is not significant.
Table~\ref{tab:tabCorrelationSpearman} shows a positive but insignificant correlation between \dis and \jerk in D--J. Since \dis is minimized and \jerk is maximized, a trade-off exists between them. The results, with a significant difference on \dis but only a small effect on \jerk, suggest that the two methods exhibit different optimization preferences under this trade-off, rather than a clear advantage on both objectives.
The reasons for the significant advantages observed in other combinations have been discussed earlier and are not repeated here. 
Overall, neither \morl nor \sorl exhibits a consistent advantage in terms of the severity of violation scenarios across \jerk-involved combinations, and their performance varies depending on the specific objective pairing.

For \textbf{objective combinations involving \sd}, the $\#SV_{x1}$ and $\#SV_{x2}$ results in Figure~\ref{fig:RQ2} show clear combination-dependent patterns. \morl achieves higher $\#SV_{S}$ values in the D--S and R--S combinations, whereas \sorl performs better in the T--S and J--S combinations. However, Table~\ref{tab:tabStatisticRQ2} indicates that only the difference in T--S is statistically significant, while the others are either insignificant or associated with small effect sizes, suggesting overall comparable performance.
Considering both objectives, \morl outperforms \sorl on both $\#SV$ metrics in D--S and R--S. In T--S, only \sd is valid, where \sorl shows a significant advantage. In J--S, \morl performs better on $\#SV_{J}$, while \sorl achieves slightly higher $\#SV_{S}$. Notably, only $\#SV_{S}$ in T--S and $\#SV_{J}$ in J--S exhibit statistically significant differences with medium effect sizes.
For the T--S combination, the strict constraint on \ttc limits its optimization space, making the problem effectively focused on \sd, under which \sorl performs significantly better.
For the J--S combination, it is consistent with the previously identified trade-off between \jerk and \sd, where \morl tends to prioritize \jerk.
Overall, the number of \sd-violation scenarios generated by the two methods varies across objective combinations, with no consistent advantage observed.
For the $SVS_{x1}$ and $SVS_{x2}$ results involving \sd, different patterns are also observed across combinations.
In D--S, \sorl outperforms \morl on both objectives, with statistically significant differences. Table~\ref{tab:tabCorrelationSpearman} shows a slight negative correlation between \dis and \sd, suggesting a certain degree of alignment between their optimization directions and reduced conflict. This relatively low-conflict setting may be one possible factor contributing to the observed advantage of \sorl.
In T--S, the presence of a strictly constrained \ttc objective limits the optimization space, making the problem closer to a single-objective setting, under which \sorl achieves a significant advantage on \sd.
In R--S, \sorl performs better on \rc, whereas \morl achieves better results on \sd, with a statistically significant but small effect size.
In J--S, \morl significantly outperforms \sorl on \jerk, while no significant difference is observed on \sd, reflecting the previously identified trade-off between \jerk and \sd.
Overall, neither \morl nor \sorl consistently generates more severe \sd-related violations across all combinations, and the severity of violations depends on the specific objective combination.

For \textbf{objective combinations involving \ttc}, valid results are only observed for the paired objectives. As discussed earlier, the optimization of \ttc is strictly constrained, which limits its impact on the overall optimization process. As a result, the observed performance differences are primarily reflected in the other objectives, whose performance has been analyzed previously.

\vspace{3pt}
\noindent \textit{\textbf{Analysis Regarding Multi-requirement Violation Metrics}}.
Due to incomputable results for \ttc-involved objective combinations in Figure~\ref{fig:RQ2}, all combinations containing \ttc are excluded from the following analysis.
For the \mv metric, \sorl outperforms \morl only in the D--J combination, while \morl performs better in all other combinations. However, Table~\ref{tab:tabStatisticRQ2} shows that none of these differences reach a medium effect size, indicating overall comparable performance, with \morl showing a slight advantage in generating a larger number of test scenarios that violate multiple requirements simultaneously.
For the \mvs metric, \sorl outperforms \morl on both objectives in the D--S combination, whereas \morl performs better on both objectives in J--S. In the remaining combinations, each method performs better on one objective. This means that the relative effectiveness of the two methods in generating more severe multi-objective violations depends on the specific objective combination, and neither method consistently outperforms the other.
Further considering the statistical results, only the D--S combination shows statistically significant differences with medium effect sizes on both objectives. Additionally, medium effect size significant differences are observed only for \rc in D--R, \dis in D--J, and \jerk in J--S. Among these, \sorl performs better in D--R and D--J, while \morl performs better in J--S. This further suggests that the severity of multi-objective violations generated by the two methods varies across combinations, although \sorl tends to produce more severe violations in some cases.
Since the multi-objective violation metrics can be regarded as a subset of the single-objective ones, the underlying reasons for the observed performance differences have already been analyzed in the single-objective violation metric analysis and are not repeated here.

\vspace{3pt}
\noindent \textit{\textbf{Analysis Regarding Collision Metric}}.
For the collision count metric $\#C$, Figure~\ref{fig:RQ2} shows that neither \sorl nor \morl consistently outperforms the other across all objective combinations, and their performance varies depending on the specific combination.
The statistical results in Table~\ref{tab:tabStatisticRQ2} further indicate that only the R--S and J--S combinations exhibit statistically significant differences with large effect sizes, while the remaining combinations are either insignificant or associated with small effect sizes, suggesting overall comparable performance.
Note that \morl outperforms \sorl in both R--S and J--S, which are also the only combinations showing significant correlations between objectives (shown in Table~\ref{tab:tabCorrelationSpearman}), and both involve the safety-related objective \sd. This suggests that the generation of collision scenarios is influenced not only by individual objectives but also by their relationships, rather than being solely driven by specific objectives, as such performance advantages are not consistently observed for other safety-related objectives (i.e., \dis and \ttc). In this context, when strong relationships between objectives (e.g., correlation or trade-offs) exist, particularly in combinations involving safety-related objectives such as \sd, \morl may more effectively leverage them in ways that increase the likelihood of generating collision scenarios.

\begin{figure}[]
    \centering
    \includegraphics[width=0.85\linewidth]{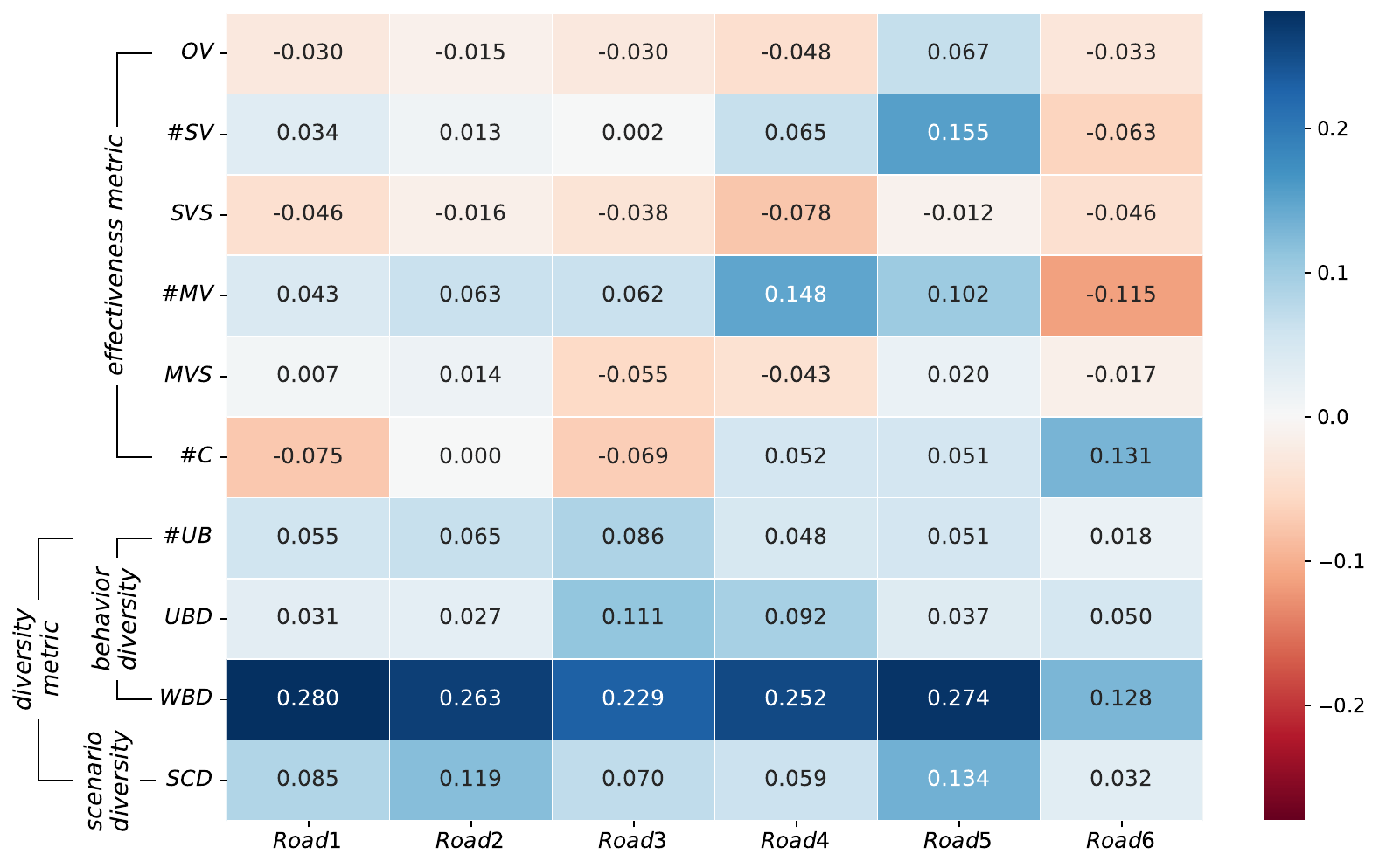}
    \caption{Heatmap of normalized average performance differences between \morl and \sorl across metrics and road structures. \ov: Objective Value; \sv: Single-requirement Violation Count; \svs: Single-requirement Violation Severity; \mv: Multi-requirement Violation Count; \mvs: Multi-requirement Violation Severity; \cc: Collision Count; \ub: Unique Behavior Count; \ubd: Unique Behavior Diversity; \wbd: Weighted Behavior Diversity; \scd: Scenario Diversity. Blue indicates \morl outperforms \sorl; Red indicates \sorl performs better. -- \textit{RQ3}}
    \label{fig:RQ3}
    \vspace{-10pt}
\end{figure}

\vspace{2pt}
\begin{center}
\fcolorbox{black}{gray!10}{
\parbox{0.97\columnwidth}{
\textbf{Conclusion for \textit{RQ2}}:
The comparative performance of \morl and \sorl is strongly dependent on the objective combination. While \morl consistently produces more diverse scenarios, neither method shows a clear overall advantage across effectiveness metrics. Instead, \sorl tends to perform better in \ttc- and \dis-involved combinations, whereas \morl shows advantages in combinations involving \rc. For \jerk and \sd, the relative performance varies across combinations. In terms of violation metrics, \morl generally generates more violation scenarios, while \sorl tends to produce more severe violations. For collision scenario generation, although no overall significant difference is observed, \morl is more likely to exploit strong interactions among objectives, particularly in safety-related combinations, to generate more collision scenarios.
}}
\end{center}

\subsection{Results for \textit{RQ3} -- Road Structure Impact}\label{subsec:RQ3}
\noindent
\textit{RQ3} investigates how different road structures affect the relative performance of \morl and \sorl. To address this RQ, we plot the heatmap in Figure~\ref{fig:RQ3} to visually represent the relative differences between \morl and \sorl. We also count the number of times each algorithm outperforms the other across all effectiveness metrics, as shown in Table~\ref{tab:tabCountRQ3}. Additionally, we conduct a statistical analysis of \morl and \sorl results for all effectiveness metrics under different road structures, as presented in Table~\ref{tab:tabStatisticRQ3}.

\subsubsection{Overall Analysis}\label{subsubsec:RQ3overall}
Figure~\ref{fig:RQ3} presents a heatmap illustrating the performance differences between \morl and \sorl across various metrics under different road structures. 
Overall, for all road structures, \morl consistently outperforms \sorl on all diversity metrics, indicating that its ability to generate diverse test scenarios is not affected by road structures and remains superior to \sorl. Due to the limited data size, we do not conduct further statistical analysis on the diversity results.
In contrast, for the effectiveness metrics, it is difficult to conclude that either \morl or \sorl consistently outperforms the other overall. Therefore, a more detailed analysis based on different road structures is necessary.

\subsubsection{Road Structure-level Analysis}
Based on Figure~\ref{fig:RQ3}, we count the number of times one algorithm outperforms the other across all effectiveness metrics, with the results shown in Table~\ref{tab:tabCountRQ3}. For \textit{Road1} and \textit{Road4}, \morl and \sorl perform better on an equal number of effectiveness metrics. For \textit{Road2} and \textit{Road5}, \morl outperforms \sorl more frequently, whereas for \textit{Road3} and \textit{Road6}, \sorl achieves better performance more often than \morl.
These results indicate that road structures have a significant impact on the effectiveness performance of the two algorithms. 
Therefore, it is necessary to conduct a deeper analysis across individual metrics to better understand how the influence of road structures leads to these differences.

\begin{table}[]
\centering
\caption{Outperformance counts between \morl and \sorl across effectiveness metrics for each road structure. \textbf{Bold} indicates the better result. -- \textit{RQ3}}
\label{tab:tabCountRQ3}
\vspace{10pt}
\resizebox{0.6\linewidth}{!}{
\begin{tabular}{lllllll}
\toprule
\multirow{2.4}{*}{\textbf{Algorithm}} & \multicolumn{6}{l}{\textbf{Road Structure}} \\ \cmidrule(r){2-7}
 & \textbf{\textit{Road1}} & \textbf{\textit{Road2}} & \textbf{\textit{Road3}} & \textbf{\textit{Road4}} & \textbf{\textit{Road5}} & \textbf{\textit{Road6}} \\ \cmidrule(r){1-7}
\morl & \textbf{3} & \textbf{3} & 2 & \textbf{3} & \textbf{5} & 1 \\
\sorl & \textbf{3} & 2 & \textbf{4} & \textbf{3} & 1 & \textbf{5} \\ \bottomrule
\end{tabular}
}
\end{table}

\subsubsection{Metric-level Analysis}
To better understand how different road structures affect the comparative performance of \morl and \sorl, we combine the results from Figure~\ref{fig:RQ3} and Table~\ref{tab:tabStatisticRQ3}, and perform a detailed analysis at the level of individual effectiveness metrics.
However, the statistical results in Table~\ref{tab:tabStatisticRQ3} show that all observed differences are either not statistically significant or associated with only negligible or small effect sizes, implying a limited overall impact of road structures on algorithm performance. Nevertheless, we further examine individual metrics to explore potential trends and better understand these subtle variations.

\begin{table}[]
\centering
\caption{Statistical results comparing \morl and \sorl across road structures and effectiveness metrics. \ov: Objective Value; \sv: Single-requirement Violation Count; \svs: Single-requirement Violation Severity; \mv: Multi-requirement Violation Count; \mvs: Multi-requirement Violation Severity; \cc: Collision Count. $e$ represents the effect size. ``NA" means not available. \textbf{Bold} indicates a moderate or larger effect size. -- \textit{RQ3}}
\label{tab:tabStatisticRQ3}
\vspace{10pt}
\resizebox{0.9\linewidth}{!}{
\begin{tabular}{ccccccccccccc}
\toprule
\multirow{2.5}{*}{\textbf{Metric}} & \multicolumn{2}{c}{\textbf{\textit{Road1}}} & \multicolumn{2}{c}{\textbf{\textit{Road2}}} & \multicolumn{2}{c}{\textbf{\textit{Road3}}} & \multicolumn{2}{c}{\textbf{\textit{Road4}}} & \multicolumn{2}{c}{\textbf{\textit{Road5}}} & \multicolumn{2}{c}{\textbf{\textit{Road6}}} \\ \cmidrule(r){2-13}
 & $\bm{p}$ & $\bm{e}$ & $\bm{p}$ & $\bm{e}$ & $\bm{p}$ & $\bm{e}$ & $\bm{p}$ & $\bm{e}$ & $\bm{p}$ & $\bm{e}$ & $\bm{p}$ & $\bm{e}$ \\ \cmidrule(r){1-13}
$\bm{OV}$ & $\ge$0.05 & 0.488 & \textless{}0.05 & 0.479 & \textless{}0.01 & 0.467 & \textless{}0.01 & 0.440 & \textless{}0.01 & 0.558 & \textless{}0.01 & 0.440 \\
$\bm{\#SV}$ & $\ge$0.05 & 0.485 & $\ge$0.05 & 0.485 & $\ge$0.05 & 0.411 & $\ge$0.05 & 0.479 & $\ge$0.05 & 0.625 & $\ge$0.05 & 0.438 \\
$\bm{SVS}$ & \textless{}0.01 & 0.449 & $\ge$0.05 & 0.478 & \textless{}0.01 & 0.433 & \textless{}0.01 & 0.376 & \textless{}0.01 & 0.443 & \textless{}0.01 & 0.441 \\
$\bm{\#MV}$ & $\ge$0.05 & 1.000 & $\ge$0.05 & 2.002 & $\ge$0.05 & 1.000 & $\ge$0.05 & 2.004 & $\ge$0.05 & 2.002 & $\ge$0.05 & 0.799 \\
$\bm{MVS}$ & \textless{}0.05 & 0.563 & $\ge$0.05 & 0.509 & \textless{}0.01 & 0.387 & \textless{}0.01 & 0.429 & $\ge$0.05 & 0.522 & $\ge$0.05 & 0.483 \\
$\bm{\#C}$ & $\ge$0.05 & 0.832 & $\ge$0.05 & NA & $\ge$0.05 & 0.832 & $\ge$0.05 & 1.502 & $\ge$0.05 & 1.000 & $\ge$0.05 & 1.251 \\ \bottomrule
\end{tabular}
}
\end{table}

\vspace{3pt}
\noindent \textit{\textbf{Analysis Regarding Objective Value Metric}}.
For the \ov results shown in Figure~\ref{fig:RQ3}, \morl outperforms \sorl only on \textit{Road5}, while \sorl achieves better performance on all other road structures.
According to the descriptions in Section~\ref{subsubsec:roads}, \textit{Road1} to \textit{Road4} correspond to straight road segments with different lane configurations (including two-lane or four-lane scenarios with either same or opposite direction traffic), while \textit{Road6} extends a two-lane straight road by introducing road damage. In contrast, \textit{Road5} is the only scenario that includes a typical four-way intersection.
Compared to the primarily straight road structures, \textit{Road5} involves more complex traffic organization and interaction patterns. The relative advantage of \morl in this scenario may suggest that it has stronger exploration capabilities in more complex road structures, which could enable it to generate more effective test scenarios. However, since the observed differences in Table~\ref{tab:tabStatisticRQ3} are not practically significant, these findings should be interpreted with caution.

\vspace{3pt}
\noindent \textit{\textbf{Analysis Regarding Single-requirement Violation Metrics}}.
For the \sv metric, \morl outperforms \sorl on all road structures except \textit{Road6}, generating a larger number of scenarios that violate a given requirement, while \sorl shows better performance on \textit{Road6}. Compared to other road structures, \textit{Road6} is characterized by a segment with poor road conditions. The observation may suggest that, in such environments, \sorl is more likely to generate more scenarios that violate a single requirement, whereas \morl tends to produce more violation scenarios in other types of road structures. However, these differences are relatively minor (Table~\ref{tab:tabStatisticRQ3}).
For the \svs metric, \sorl consistently outperforms \morl across all road structures, indicating that the scenarios it generates tend to exhibit higher degrees of violation with respect to a given requirement, regardless of road conditions. However, the overall differences in Table~\ref{tab:tabStatisticRQ3} are still relatively small.

\vspace{3pt}
\noindent \textit{\textbf{Analysis Regarding Multi-requirement Violation Metrics}}.
For the \mv metric, results similar to those observed for the \sv metric are found: \sorl outperforms \morl on \textit{Road6}, generating more test scenarios that simultaneously violate multiple requirements, while \morl performs better on the other road structures. For the \mvs metric, the effects of different road structures on algorithm performance are more variable, and no algorithm consistently generates more severe violations across all road structures. Overall, the differences between the algorithms in Table~\ref{tab:tabStatisticRQ3} are relatively small, while their performance still varies across road structures, suggesting that it could be considered under specific road conditions.

\vspace{3pt}
\noindent \textit{\textbf{Analysis Regarding Collision Metric}}.
For the \cc metric, no difference is observed between the two algorithms on \textit{Road2}. A further examination reveals that neither algorithm generates collision scenarios under this road structure. On \textit{Road1} and \textit{Road3}, \sorl produces more collision scenarios, whereas \morl generates more collision scenarios on \textit{Road4} to \textit{Road6}. Although the differences between the two algorithms are relatively small (Table~\ref{tab:tabStatisticRQ3}), the results suggest that road structure may have an influence on their ability to generate collision scenarios. This observation may be related to differences in traffic interactions and driving complexity across road structures.

\vspace{-5pt}
\begin{center}
\fcolorbox{black}{gray!10}{
\parbox{0.97\columnwidth}{
\textbf{Conclusion for \textit{RQ3}}:
\morl consistently outperforms \sorl on all diversity metrics across different road structures, demonstrating its robustness in generating diverse test scenarios. For effectiveness metrics, neither algorithm exhibits a consistent overall advantage, and their relative performance varies across road structures. Although variations can be observed at both the road structure and metric levels, most differences are either not statistically significant or associated with negligible or small effect sizes, indicating limited practical impact. Consequently, road structures do not fundamentally change the overall effectiveness of the algorithms, but still influence their performance under specific conditions.
}}
\end{center}

\subsection{Discussion}

\noindent \textit{\textbf{The trade-off between diversity and effectiveness}}.
Across all RQs, \morl consistently demonstrates an advantage in exploration, enabling broader coverage and generating a larger number of diverse violation scenarios, as reflected by both diversity metrics and violation count-based metrics (i.e., \sv and \mv). In contrast, \sorl may produce stronger violations under specific objective combinations or road structures, as observed in certain effectiveness metrics related to violation severity. This reflects an inherent trade-off between exploration and optimization: while \morl generally enhances scenario diversity and coverage, \sorl occasionally concentrates its efforts on achieving more severe violations. Although such differences do not consistently show clear advantages across all metrics, they highlight that the two approaches prioritize different aspects of scenario generation.

\vspace{3pt}
\noindent \textit{\textbf{Metric-level variability and interdependence}}.
The evaluation metrics in this study capture different aspects of scenario quality, including overall performance (\ov), single-requirement violations (\sv and \svs), multi-requirement violations (\mv and \mvs), collision occurrences (\cc), behavioral diversity (\ub, \ubd, and \wbd), and scenario diversity (\scd). Some metrics are inherently related, for example, multi-requirement violation metrics (\mv and \mvs) are computed based on scenarios that already violate a single requirement (\sv and \svs). Despite this dependency, single- and multi-requirement metrics provide complementary perspectives: single-requirement metrics focus on violations of individual objectives, while multi-requirement metrics highlight interactions and simultaneous violations across multiple objectives. Similarly, frequency-based metrics (\sv, \mv, and \cc) capture how often violations occur, whereas severity-based metrics (\svs and \mvs) quantify the magnitude of these violations. Behavior and scenario diversity metrics (\ub, \ubd, \wbd, and \scd) assess different dimensions of agent behavior and scenario coverage. These metrics together provide a comprehensive and complementary view of scenario quality, allowing subtle variations in algorithm performance to be captured across multiple perspectives.

\vspace{3pt}
\noindent \textit{\textbf{Implications of objective categories on method effectiveness}}.
As discussed in Section~\ref{subsubsec: requirements_objective_functions}, the objectives in this study can be broadly categorized into safety and functionality requirements.
However, the results do not reveal a consistent advantage of either \morl or \sorl at this categorical level. Instead, the comparative performance of the two methods is highly dependent on specific objective combinations and their interactions.
For instance, \sorl tends to perform better in combinations where one objective has a constrained optimization space (e.g., involving \ttc), making the problem closer to single-objective optimization, whereas \morl shows advantages in combinations where stronger relationships or interactions between objectives exist.
These observations suggest that the effectiveness of multi-objective test scenarios is not well explained by high-level objective categories, but rather by specific objective combinations and their interactions. As a result, the choice between \morl and \sorl should be guided by these factors in the given testing context, rather than relying solely on high-level objective categories such as safety or functionality.

\vspace{3pt}
\noindent \textit{\textbf{Guidelines for algorithm selection}}.
Based on the above observations, the selection between \morl and \sorl should be guided by the specific testing objectives, scenario characteristics, and road conditions. 
When maximizing scenario diversity and coverage is the primary goal, \morl is generally more suitable due to its consistent advantage in exploration and its ability to generate a larger number of diverse violation scenarios. In contrast, when the focus is on generating scenarios with more severe violations, \sorl may be preferred, as it tends to concentrate optimization on producing higher violation severity under certain conditions.
More specifically, the relative effectiveness of the two methods depends strongly on the objective combination. For example, \sorl tends to perform better in combinations involving more constrained objectives (e.g., \ttc), where the problem becomes closer to single-objective optimization, whereas \morl shows advantages in combinations with stronger interactions among objectives (e.g., involving \sd), where coordinated optimization is required. 
Additionally, neither method shows a consistent advantage across all effectiveness metrics, and their performance may vary under specific road conditions, particularly in more complex scenarios or under poor road settings. Therefore, selecting an appropriate approach requires jointly considering the testing goals, the characteristics of objective combinations, and the scenario environment.

\section{Threats to Validity}\label{sec:threats}
\noindent 
\textit{Conclusion validity} is concerned with the reliability of the conclusion drawn from the empirical study, which might be influenced by the inherent randomness of the simulator. We mitigate this threat by repeating each execution 100 times and using appropriate statistical tests following well-known guidelines~\cite{arcuri2011practical}.
\textit{Construct validity} relates to the evaluation metrics. For a fair comparison, we design six effectiveness metrics and four diversity metrics for AV testing, and apply the same metric to evaluate the performance of both \sorl and \morl. 
\textit{Internal validity} concerns with the parameter settings. To mitigate the potential threats to internal validity, we determined the hyperparameters by employing an automated hyperparameter optimization framework named Optuna~\cite{akiba2019optuna}. Besides, we acknowledge that the performance of RL algorithms might be improved by tuning agent settings, e.g., formulating the state space using sensor data such as camera images, and designing action space covering more environmental objects. Thus, we will investigate other \sorl and \morl designs in the future. 
\textit{External validity} is about the generalizability of the empirical evaluation. We used one subject system and one simulator to conduct the experiment, which could potentially threaten the external validity. To mitigate this issue, we chose the advanced end-to-end AV controller (i.e., Interfuser) and the high-fidelity simulator (i.e., CARLA). However, we acknowledge that conducting experiments with more case study systems could strengthen our conclusions. Notably, \sorl and \morl can be easily adapted to test other AV systems with minimal effort.
Moreover, focusing on a specific set of objective combinations and road structures might limit the generalizability of our findings. To minimize this threat, we systematically formulate ten distinct combinations by pairwise pairing five fundamental driving requirements encompassing both safety and functional aspects, and conduct our evaluations across six representative road structures provided by CARLA Leaderboard~\cite{carlaleaderboard}.

\section{Related Work}\label{sec:related_work}
\noindent Testing AVs in various driving scenarios is critical in ensuring their safety and functionality. However, due to the complexity of the AV operating environments, there is theoretically an infinite number of possible driving scenarios. Therefore, we need to identify critical scenarios that could violate safety or functional requirements.
Search-based methods are widely adopted to generate critical scenarios under the guidance of fitness functions~\cite{zhong2022neural,10645815,10234383,zhou2023specification,huai2023scenorita}. For example, 
Li et al.~\cite{li2020av} proposed AV-Fuzzer, an approach integrating fuzz testing with evolutionary search, to identify critical scenarios that cause safety violations. Haq et al.~\cite{haq2022efficient} proposed SAMOTA, an efficient AV testing approach that combines multi-objective search with surrogate models. SAMOTA employs multi-objective search to generate critical scenarios and improves testing efficiency by simulating the simulator with surrogate models. To test AV against traffic laws, Sun et al.~\cite{sun2022lawbreaker} proposed LawBreaker, which adopts a fuzzing algorithm to search for scenarios that can effectively violate traffic laws.
Moreover, Huai et al.~\cite{huai2023scenorita} proposed scenoRITA, a test generation approach that utilizes an evolutionary algorithm equipped with a novel gene representation and multiple test oracles to automatically generate diverse, fully mutable critical scenarios that effectively reveal ADS violations.
Though search-based approaches have shown promising performance, they have limited effectiveness in handling tasks requiring runtime sequential interactions with the environment, which are essential when manipulating dynamic objects and ensuring precise control over them. Another challenge is that the AV operating environment is dynamically changing and uncertain, while search-based approaches cannot effectively adapt to environmental changes.

RL-based approaches generate critical scenarios through intelligent agents that interact with and configure the AV operating environment~\cite{chen2021adversarial,lu2022learning,10172658,feng2023dense,giamattei2025reinforcement}. For example, 
Lu et al.~\cite{lu2022learning} proposed DeepCollision, an RL-based environment configuration framework that generates scenarios to uncover collisions of AV. To test lane-changing models, Chen et al.~\cite{chen2021adversarial} proposed an RL-based adaptive testing framework to generate time-sequential adversarial scenarios. Similarly, Doreste et al.~\cite{doreste2024adversarial} proposed an adversarial method that models NPCs with independent RL-based agents to challenge the ADS under test. Feng et al.~\cite{feng2023dense} proposed a dense RL-based approach to learn critical scenarios from naturalistic driving data by editing the Markov decision process to eliminate non-safety-critical states. Wei et al.~\cite{wei2024interactive} proposed an interactive critical scenario generation method that utilizes real-world in-depth crash data combined with generative adversarial networks (CTGAN) and a reinforcement learning algorithm (TD3) to dynamically generate highly challenging scenarios for assessing AV safety.
To test multiple requirements, Haq et al.~\cite{10172658} proposed MORLOT that adapts single-objective RL and multi-objective search to generate test suites to violate as many requirements as possible. Notice that MORLOT is designed to violate multiple \textit{independent} requirements by developing a set of scenarios, each violating one independent requirement. However, in reality, AV requirements are often interdependent and must be evaluated simultaneously.
For example, failing to meet a functional requirement (e.g., not completing a route) may violate a safety requirement (e.g., causing a crash), highlighting the need for approaches simultaneously evaluating multiple requirements. 

Different from the above works, we conduct an empirical comparison of multi-objective and single-objective RL approaches, considering interdependent requirements under different objective combinations and road structures. Our study provides insights into their performance differences and offers preliminary guidelines for selecting appropriate RL strategies in AV testing.

\section{Conclusion and Future Work}\label{sec:conclusion}
\noindent
In practice, many Autonomous Vehicle (AV) requirements are inherently interdependent and must be evaluated simultaneously to ensure comprehensive testing. To this end, we present a systematic comparison between \morl and \sorl for test scenario generation, considering multiple evaluation metrics, objective combinations, and road structures.
The results indicate that, while the two approaches achieve comparable performance in terms of effectiveness, they differ in how requirement violations are reflected. Specifically, \morl tends to generate a larger number of violation scenarios, whereas \sorl is more likely to produce violations with higher severity under certain conditions. Further analysis shows that these differences are influenced by specific objective combinations and, to a lesser extent, by road structures. Rather than demonstrating a clear overall superiority, the two approaches emphasize different aspects of scenario generation, reflecting a trade-off between exploration and focused optimization. These findings highlight the importance of aligning the choice of method with the desired testing objectives and environmental conditions.
Furthermore, \morl demonstrates a clear and consistent advantage in diversity, generating a broader range of test scenarios at both the behavior and scenario levels.
The replication package is provided in an online repository~\cite{MORLrepo}. 

Future work can extend this study in several directions. First, while this work considers a representative set of objectives and road configurations, exploring a broader range of scenarios and objective formulations could further validate the generality of the findings. Second, a deeper analysis of the interaction between objectives and learning dynamics may provide more detailed insights into why \morl and \sorl exhibit different behaviors under varying conditions, and support the development of more refined guidelines for algorithm selection. Finally, investigating hybrid or adaptive approaches that balance exploration and optimization could help leverage the complementary strengths of the two algorithms.

\section*{Acknowledgments}
This work is supported by the Co-tester project (No. 314544) and the Co-evolver project (No. 286898/F20), funded by the Research Council of Norway.
The work is also partially supported by the RoboSAPIENS project funded by the European Commission’s Horizon Europe programme under grant agreement number 101133807.
Aitor Arrieta is part of the Software and Systems Engineering research group of Mondragon Unibertsitatea (IT1519-22), supported by the Department of Education, Universities and Research of the Basque Country. Aitor Arrieta is also supported by the Spanish Ministry of Science, Innovation and Universities (project PID2023-152979OA-I00), funded by MCIU /AEI /10.13039/501100011033 / FEDER, UE.

\bibliographystyle{unsrt}  
\bibliography{references}

\end{document}